\documentclass{aa}  

\usepackage{natbib}
\bibpunct{(}{)}{;}{a}{}{,} 
\usepackage{graphicx}
\usepackage{setspace}
\usepackage{amsmath}

\usepackage{eucal}
\usepackage[colorlinks=True,allcolors=blue]{hyperref}
\usepackage{txfonts}
\usepackage{caption}
\usepackage{subcaption}
\usepackage{multirow}
\usepackage{lscape}
\usepackage{longtable}
\usepackage{rotating}
\begin{document}

   \title{A statistical study of ionized outflows in low-mass galaxies from DESI DR1}
   \authorrunning{M. Romano et al.}
   \titlerunning{Ionized outflows in low-mass galaxies with DESI}

\author{M. Romano\thanks{E-mail: mromano@mpifr-bonn.mpg.de}\inst{1,2}
\and
A. Saintonge\inst{1,3}
\and
C. Circosta\inst{4}
\and
C. Cicone\inst{5}
\and
D. Donevski\inst{6,7,8}
\and
D. Scholte\inst{9}
\and
B. Hagedorn\inst{5}
}

\institute{
Max-Planck-Institut f\"ur Radioastronomie, Auf dem H\"ugel 69, 53121 Bonn, Germany
\and
INAF - Osservatorio Astronomico di Padova, Vicolo dell'Osservatorio 5, I-35122, Padova, Italy
\and
Department of Physics and Astronomy, University College London, London, WC1E 6BT, UK
\and
Institut de Radioastronomie Millimétrique (IRAM), 300 rue de la Piscine, 38400 Saint-Martin-d’Hères, France
\and
Institute of Theoretical Astrophysics, University of Oslo, P.O Box 1029, Blindern, 0315 Oslo, Norway
\and
National Center for Nuclear Research, Pasteura 7, 02-093 Warsaw, Poland
\and
SISSA, Via Bonomea 265, 34136 Trieste, Italy
\and
INAF - Osservatorio Astronomico di Trieste, via Tiepolo 11, I-34131, Trieste, Italy
\and
Institute for Astronomy, University of Edinburgh, Royal Observatory, Blackford Hill, Edinburgh EH9 3HJ, UK
}


 
\abstract
{We present a statistical analysis of ionized gas outflows in a large sample of low-mass ($\mathrm{log(M_{*}/M_{\odot})\lesssim10}$) galaxies at $z<0.45$ from the first data release of the Dark Energy Spectroscopic Instrument survey (DESI DR1). Using [O~III] and H$\alpha$ emission lines, we identify $\sim5800$ dwarf galaxies exhibiting robust signatures of ionized outflows. Most are star-forming systems located above the star-forming main sequence, with a smaller fraction of composites and AGN hosts. The derived mass-loading factors span $\eta = 0.001-1$, with a median $\eta\sim0.01$, confirming previous findings from smaller datasets at similar stellar masses. These values are $1$--$2~$dex lower than predicted by cosmological simulations, suggesting that the ionized phase traces only a minor fraction of the total outflow mass budget. Outflow velocities are typically sufficient to enrich the circumgalactic medium in $\gtrsim80\%$ of cases, while escape into the intergalactic medium appears efficient mainly in composite- and AGN-host galaxies. We find that the mass-loading factor decreases with stellar mass, broadly reproducing the trends predicted by simulations despite the strong normalization offset. By comparing individual outflow detections with stacked spectra of the general low-mass star-forming galaxy population from DESI, we show that mass-loading factors inferred from the stacks are systematically lower than those of individual detections, but the difference decreases toward larger offsets above the star-forming main sequence. Together with a positive correlation between mass-loading factor and specific star-formation rate, this suggests that enhanced star formation and burstiness play an important role in regulating feedback efficiency and shaping galactic winds. Our results indicate that ionized gas outflows play a significant role in regulating star formation and the baryon cycle of low-mass galaxies, but also emphasize the need to extend these studies to multiple gas phases to achieve a complete understanding of feedback in galaxy evolution.} 
  

\maketitle
%

\section{Introduction}
Stellar and active galactic nuclei (AGNs) feedback are considered a fundamental component of galaxies' evolution. In fact, they play a leading role in regulating their baryon cycle by producing large-scale winds able to heat or remove a significant amount of gas from the interstellar medium (ISM), possibly polluting the surrounding circum- and intergalactic medium (CGM and IGM) with metal-enriched material (see e.g., \citealt{Peroux20,Tacconi20}). Their ability to regulate or completely stop star formation is required by cosmological simulations to match both the low- and high-mass ends of the predicted halo mass function to the observed luminosity function of galaxies, thought to be primarily shaped by star-formation and AGN feedback, respectively (e.g., \citealt{Silk12,Behroozi13,Pillepich18}). At the same time, they are needed to explain a number of fundamental scaling relations (e.g., \citealt{Mannucci09,McGaugh12,Somerville15,Curti20}), or the observed black hole-galaxy co-evolution (e.g., \citealt{Kormendy13,King15}).  

Nearby low-mass galaxies (with stellar mass lower than $10^{10}~M_{\rm{\odot}}$; e.g., \citealt{McCormick18,Marasco23}) are particularly interesting for investigating outflow-driven processes. Their shallow gravitational potential wells allow galactic winds to efficiently expel gas and dust out of their stellar disk (e.g., \citealt{Christensen18,McCormick18,Romano23,Romano24}). Additionally, their low metallicities and compact nature make them excellent analogs of distant galaxies, offering valuable insight into mechanisms shaping galaxy evolution in the high-redshift Universe (e.g., \citealt{Izotov21,Shivaei22}). The general theoretical consensus is that stellar feedback dominates in low-mass galaxies, with high-mass stars exploding as supernovae and injecting energy and momentum in the ISM, accelerating the surrounding gas and producing galactic-scale outflows (e.g., \citealt{Hopkins12,Muratov15,Hayward17,Nelson19}). This is also supported by observational studies which detect star-formation driven outflows in dwarf starbursting galaxies (e.g., \citealt{Strickland00,Cicone16, Martin-Navarro18,Romano23,Carniani24,Hagedorn26}). On the other hand, over the past decade, evidence for AGN activity in nearby dwarf galaxies has been growing, as witnessed by thousands of detections via optical and infrared tracers (e.g., \citealt{Reines13,Sartori15,Kaviraj19,Manzano19,Pucha25,Rodriguez25,Rodriguez26}). This challenges the current paradigm for which AGN feedback is not an important channel of star formation regulation in low-mass galaxies, and is being further theoretically investigated (e.g., \citealt{Dashyan18,Trebitsch18,Koudmani19,Arjona24}).

From the observational point of view, outflows have been detected in different kinds of galaxies and across diverse cosmic epochs. Given their multi-phase nature (e.g., \citealt{Cicone18,Fluetsch19}), different techniques, instruments, and gas tracers have been adopted for their characterization, covering the whole electromagnetic spectrum from the rest-frame UV/optical and far-infrared (FIR) emission (for the warm, ionized, and cold atomic and molecular gas; e.g., \citealt{Erb12,Heckman15,Cicone16,Roberts-Borsani20,Calabro22,Concas22,Hagedorn26}), to the X-ray emission (for the more hot phase; e.g., \citealt{Ott05,Tombesi15,McQuinn18}). Evidence for outflowing gas are typically traced by blueshifted UV absorption and FIR lines (e.g., \citealt{Erb12,Veilleux13,Talia17,Riechers21}), as well as in emission by searching for broad wings in the observed line spectra (e.g., \citealt{Arribas14,Janssen16,Herrera19,Marasco23,Romano23}), or by spectral and photometric disk-bulge/CGM decomposition (e.g., \citealt{McCormick18,Gallagher19,Concas22}). 

Several works have taken advantage of these methods to investigate the incidence and strength of outflows in low-mass galaxies, now accessible up to the Epoch of Reionization thanks to JWST (e.g., \citealt{Carniani24,Matthee24,Ivey25,Saldana25,Xu25}). However, no consensus has been reached yet regarding the importance of (stellar or AGN) feedback in these galaxies. For instance, \cite{Marasco23} studied the properties of the warm, ionized gas in outflows in a sample of 19 dwarf, starburst galaxies with $\mathrm{log(M_{*}/M_{\odot})\sim7-10}$ using H$\alpha$ emission from MUSE data. They found typical mass-loading factors (namely the outflow efficiency, defined as the ratio between the mass outflow and star-formation rates) of the order of $\sim0.02$, suggesting that the ionized gas entrained by the winds will stay bound to the galaxies, eventually enriching the CGM and falling back in a galactic fountain cycle, but not escaping the dark matter halos. \cite{Hagedorn26} used medium-resolution single-slit spectroscopy to investigate ionized outflows in low-mass galaxies with $\log(M_\ast/M_\odot)\sim8.5-10$, detecting outflow signatures in $\sim30\%$ of their targets through asymmetric and blueshifted emission-line wings that were not identified in the lower-resolution SDSS spectra. They found a higher incidence of outflows in galaxies with enhanced SFRs and a mean mass-loading factor of $\eta\sim0.07$, broadly consistent with the low ionized mass-loading factors found by \cite{Marasco23}. Ionized outflows have also been investigated in AGN-host dwarf galaxies. \cite{Rodriguez25} studied 13 nearby, AGN-host dwarfs ($\mathrm{log(M_{*}/M_{\odot})\sim9-10}$) from MaNGA \citep{Bundy15} finding relatively low ionized mass-loading factors and suggesting that most of the outflowing gas remains gravitationally bound to the galaxies. More recently, using a much larger sample of dwarf AGNs from DESI DR1, \cite{Rodriguez26} found that their ionized outflows can reach velocities sufficient to overcome the gravitational potential of their hosts, suggesting that AGN feedback may play an important role in redistributing gas in low-mass galaxies. On the other hand, \cite{McQuinn19} used deep H$\alpha$ observations to characterize the ionized outflows in 12 nearby, starburst dwarfs ($\mathrm{log(M_{*}/M_{\odot})\sim7.0-9.3}$), finding mass-loading factors in the range $0.2-7$, more than a factor $\sim100$ larger than \cite{Marasco23}, but still concluding that most of the expelled material will stay bound to the galaxies' halos. \cite{Manzano19} used Keck spectroscopy to fit the [O~III]$\lambda\lambda4959,5007$ doublet in 13 dwarf (AGN and star-forming) galaxies with stellar masses $\mathrm{log(M_{*}/M_{\odot})\sim9-10}$ selected from SDSS, with all of them showing outflow velocities larger than the escape velocities, thus pushing the gas into the IGM. Similarly, \cite{Romano23} analyzed the atomic phase of the outflow in a sample of local dwarf sources ($\mathrm{log(M_{*}/M_{\odot})\sim5-10}$) by looking at their [C~II]$\lambda158\mu\rm{m}$ emission lines from \textit{Herschel}, some of them being part of the sample studied by \cite{Marasco23}. They found that outflows can typically escape the gravitational potential of their hosts, and mass-loading factors of order of unity, $\sim2$ orders of magnitude larger than what reported in previous studies analyzing the ionized gas phase, thus suggesting that different gas phases can possibly contribute in a different way to the overall mass and energy outflow budget (e.g., \citealt{Cicone18,Fluetsch19,Fluetsch21}). Studies of ionized gas at high redshift with rest-frame optical emission from JWST show a possible evolution of outflow properties with cosmic time but with a large scatter, with some works finding modest outflow velocities and typical mass-loading factors $\lesssim1$ (e.g., \citealt{Saldana25,Xu25}), and other pointing towards more efficient outflows and easier IGM enrichment (e.g., \citealt{Carniani24}). 

All these contrasting results highlight the need for larger, statistically robust studies to better constrain the incidence and efficiency of outflows in low-mass galaxies. The aim of this work is to address this gap by leveraging the unprecedented statistical power of the DESI cosmological survey to characterize ionized outflows in a large sample of nearby low-mass galaxies. In a forthcoming companion paper, we will extend this analysis to high-redshift galaxies, enabling a direct comparison of outflow properties across cosmic time.

The paper is structured as follows. In Sect.~\ref{sec:sample}, we provide a description of the data. Section~\ref{sec:method} reports the method used to identify and characterize outflows in low-mass galaxies. Section~\ref{sec:results} shows the results obtained, that are later discussed in Sect.~\ref{sec:discussion}. We summarize our findings in Sect.~\ref{sec:summary}. Throughout the paper, we adopt a $\Lambda-\rm{CDM}$ cosmology with $H_0=70~\rm{km~s^{-1}~Mpc^{-1}}$, $\Omega_{\rm{m}}=0.3$, and $\Omega_{\rm{\Lambda}}=0.7$, and assume a \cite{Chabrier03} initial mass function (IMF).

\section{Data}\label{sec:sample}
\subsection{The DESI survey}
The Dark Energy Spectroscopic Instrument (DESI)\footnote{\url{https://www.desi.lbl.gov/the-desi-survey/}.} is a 5000-fiber multiplexed spectrograph mounted on the Mayall 4-meter telescope at Kitt Peak National Observatory (KNPO). The fibers have an angular diameter of $1.5''$ \citep{Poppett24}, corresponding to physical scales ranging from a few tens of parsecs to several kiloparsecs over the redshift range of our sample (see Sect.~\ref{subsec:outflow_rate}). It covers the spectral range $3600-9800~\rm{\AA}$, with a resolving power that ranges from 2000 to 5500 from the shortest to the longest wavelength channel \citep{DESI22,Poppett24}. DESI is conducting a 8-year cosmological survey that will map a total area of $\sim17,000~\rm{deg^2}$ at the time of its completion, collecting the redshift of about 60 million galaxies and quasars from their optical spectra, with the main goal of providing a wide 3D map of the large-scale structure of the Universe and study the nature of dark energy (e.g., \citealt{Levi13,DESI16,Schlafly23}). However, its exquisite spectroscopic capabilities make it also a huge resource for extragalactic studies. In this work, we take advantage of data from DESI Data Release 1 (DR1\footnote{\url{https://data.desi.lbl.gov/doc/releases/dr1/}}; \citealt{DESI25}), released on March 2025, to investigate the impact of galactic outflows in dwarf galaxies. DESI DR1 includes more than 18 million unique targets from the Main Survey observations ($\sim13~$million of which are classified as galaxies), plus all the data taken as part of the Survey Validation from the Early Data Release \citep{DESI24}. All the details about DR1 can be found in \cite{DESI25}. For the scope of this work, we made a cross-match between the Dwarf Galaxy Catalog\footnote{\url{https://data.desi.lbl.gov/doc/releases/dr1/vac/extragalactic-dwarfs/}} and the Stellar Mass and Emission Line Catalog\footnote{\url{https://data.desi.lbl.gov/doc/releases/dr1/vac/stellar-mass-emline/}}, defining our dwarf parent sample, which is briefly described in the next Section. 

\subsection{The Dwarf Parent Sample}\label{subsec:catalog}
We started collecting dwarf galaxies from the Dwarf Galaxy Catalog as part of the DESI Value-added Catalogs released with DR1. It contains a total of $647,241$ dwarf galaxies, selected from the following primary target classes: the bright galaxy survey (BGS; \citealt{Hahn23}), emission-line galaxies (ELG; \citealt{Raichoor23}), and the LOW-Z secondary target program (SCND; \citealt{Darragh23}). All galaxies lie within the redshift range $0.001<z<0.5$. 

Stellar masses and star-formation rates (SFRs) for these galaxies were originally derived through spectral energy distribution (SED) fitting by \cite{Siudek24}. However, in this work we adopt the physical parameters from \cite{Zou24}, as provided in the general Stellar Mass and Emission Line Catalog of DESI DR1. While both studies employ the SED-fitting code CIGALE v2022.1 \citep{Boquien19} to infer physical properties, there are important differences in their input assumptions. In particular, \cite{Zou24} adopt a grid of (sub-)solar stellar metallicities, in constrast to the fixed solar metallicity used by \cite{Siudek24} for the Dwarf Galaxy Catalog. Moreover, \cite{Zou24} exclude AGN templates that could otherwise produce spurious results when the rest-frame mid-infrared coverage is insufficient, as for our galaxies (e.g., \citealt{Ciesla15}). We note, however, that neglecting AGN emission may lead to overestimated SFRs in AGN-host galaxies, depending on the AGN contribution to the SED (e.g., \citealt{Ciesla15,Florez20}). While the magnitude of this bias cannot be reliably quantified for individual sources given the available photometric coverage, it introduces an additional systematic uncertainty in the inferred SFRs of AGN-host galaxies (see Sect.~\ref{subsec:outflow_rate}).

By cross-matching the two catalogs, we identified 645,790 sources in common. Since we want to focus on the low-mass regime, we further restricted the sample to galaxies with $\mathrm{log(M_{*}/M_{\odot})}<10$. Hereafter, we refer to this mass-selected cross-matched dataset as the dwarf parent sample, adopting the stellar masses and SFRs from \cite{Zou24} throughout.

\begin{figure*}[t!]
    \begin{center}
        \includegraphics[width=0.9\textwidth]{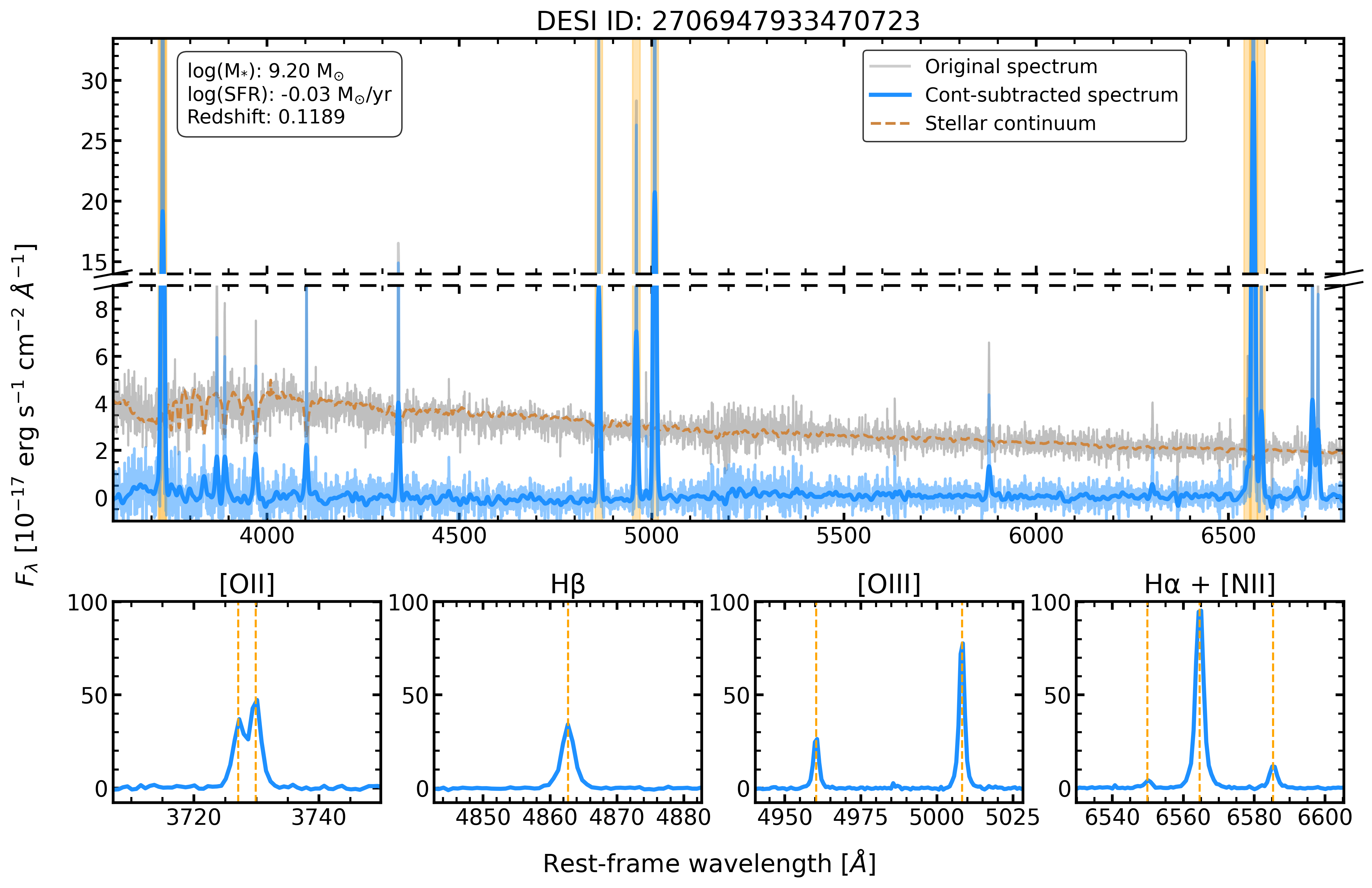}
        \end{center}
    \caption{Example of reduced spectrum for one of the dwarf galaxies in DESI. The top two panels show the spectrum at rest-frame wavelength between 3600 and 6800~$\AA$, with a broken y-axis layout for better visualization. The original spectrum is reported in gray, with overlaid the stellar continuum estimated from \texttt{FastSpecFit} as a brown, dashed line. In light-blue, the continuum-subtracted spectrum is shown. The ticker blue line displays the same spectrum, but smoothed for clarity. Physical properties and redshift of the galaxy are reported in the framed text. The bottom panels show a zoom-in on some of the strongest emission lines (highlighted in orange in the top panels) in the un-smoothed spectrum. Their rest-frame wavelengths are marked with dashed lines.}
    \label{fig:spectrum_example}
\end{figure*}

\subsection{Selection of outflow candidates}

To identify galaxies hosting candidate ionized outflows, we cross-matched the dwarf parent sample with the \texttt{EmFit} catalog\footnote{\url{https://data.desi.lbl.gov/doc/releases/dr1/vac/emfit/}} computed for sources in DR1, obtaining $636,086$ galaxies. \texttt{EmFit} is a Python-based emission-line fitting code specifically developed for DESI low-redshift sources. It uses a nonlinear least-squares fitting algorithm to fit H$\beta$, [O~III]$\lambda\lambda4959,5007$, [N~II]$\lambda\lambda6548,6583$, H$\alpha$, and [S~II]$\lambda\lambda6716,6731$. \cite{Pucha25} ran this code on all sources with good spectra, irrespective of their photometry and stellar mass measurements, with the only restriction to be at redshift $0.001\leq z \leq 0.45$, in order to exclude contamination from Galactic sources and to include all the strongest emission lines used for the fitting in the observed spectra (lower and upper redshift limits, respectively). A detailed description of the fitting procedure, uncertainties, and quality cuts, can be found in \cite{Pucha25}. For the purposes of this paper, we just specify here that the emission lines are modeled within separate spectral windows, such that the statistical preference for additional Gaussian components is evaluated independently within each fitting region, with additional components retained when the improvement in $\chi^2$ corresponds to a $>5\sigma$ confidence level. We focus on the [O~III]$\lambda5007$ emission line, as it is a good tracer of ionized outflowing gas in the optical and is less affected by contamination from nearby emission lines or AGN broad-line region emission (e.g., \citealt{Liu13,Manzano19,Singha22}). For [O~III]$\lambda5007$, the narrower Gaussian is identified as the primary component, while an additional broader component can trace either outflowing gas or more complex, double-peaked gas kinematics \citep{Pucha25}. We selected the sources for which this secondary component was classified as an outflow according to the criteria described in \cite{Pucha25}. This shrinks the sample to $7621$ candidate outflow-host sources. Notably, all the selected [O~III] emitters have H$\alpha$ lines that can be inspected as well to further constrain outflow properties. Afterwards, we ran our own fitting code on those galaxies, to i) search for outflow signatures in their spectra while having control of the fitting procedure, and ii) estimate additional spectral parameters needed to compute the outflow properties. The adopted fitting procedure is described in Sect.~\ref{sec:method}.

\section{Methods}\label{sec:method}
\subsection{Outflow identification}\label{subsec:fitting}
To properly characterize signatures of outflowing gas in our dwarf galaxies, we performed a multi-Gaussian fit to their 1D optical spectra. Our fitting procedure builds on the methodology developed by Circosta et al. (in prep.) to characterize [O~III] emission in DESI AGN, which we adapted here to the analysis of low-mass galaxies and extended to H$\alpha$. In the following, we summarize our method.

First, we got access to the DESI DR1 optical spectra through the SPectra Analysis \& Retrievable Catalog Lab (SPARCL; \citealt{Juneau24}), as managed by the NSF National Optical-Infrared Astronomy Research Laboratory. 
For each source, we corrected the spectrum for galactic reddening using the \cite{Fitzpatrick99} dust extinction law with $R_\mathrm{V}=3.1$. Then, we retrieved the stellar continuum estimated from \texttt{FastSpecFit}, a stellar continuum and emission-line modeling code optimized for DESI (\citealt{Moustakas23}). We thus subtracted the stellar continuum from each spectrum, and shifted it into the rest frame. Figure~\ref{fig:spectrum_example} shows an example of optical spectrum before and after continuum subtraction for a galaxy at $z=0.119$ with $\rm{log(M_{*}/M_{\odot})=9.20}$ and $\rm{log(SFR/M_{\odot}~yr^{-1})=-0.03}$.

We focused the search for broad emission-line components indicative of ionized outflowing gas on the brightest emission lines in the spectra, that is [O~III]$\lambda$5007 (hereafter, [O~III]) and H$\alpha$. For the [O~III], we fit the line with a single and double Gaussian model, the latter of the form:
\begin{multline}\label{eq:gauss_model}
    F_{\lambda} = \frac{F_{\rm{narrow}}}{\sigma_{\rm{narrow}} \sqrt{2\pi}}\exp{\left(\frac{-(\lambda-\lambda_{\rm{narrow}})}{2\sigma^2_{\rm{narrow}}}\right)} \\
    + \frac{F_{\rm{broad}}}{\sigma_{\rm{broad}} \sqrt{2\pi}}\exp{\left(\frac{-(\lambda-\lambda_{\rm{broad}})}{2\sigma^2_{\rm{broad}}}\right)},
\end{multline}
where $F_{\rm{narrow}}$, $\lambda_{\rm{narrow}}$, and $\sigma_{\rm{narrow}}$ correspond to the integrated flux, central wavelength, and velocity dispersion of the narrow Gaussian component, respectively, while $F_{\rm{broad}}$, $\lambda_{\rm{broad}}$, and $\sigma_{\rm{broad}}$ represent the same parameters but for the broad component. The Gaussian widths obtained from the fitting correspond to the observed line widths and are not corrected for instrumental broadening at this stage. We account for the wavelength-dependent DESI spectral resolution when deriving the intrinsic outflow kinematics, by deconvolving the instrumental dispersion from the measured line widths (see Sect.~\ref{subsec:outflow_rate}).

\begin{figure*}[t]
    \begin{subfigure}{0.5\textwidth}
    \centering
        \includegraphics[width=0.98\linewidth]{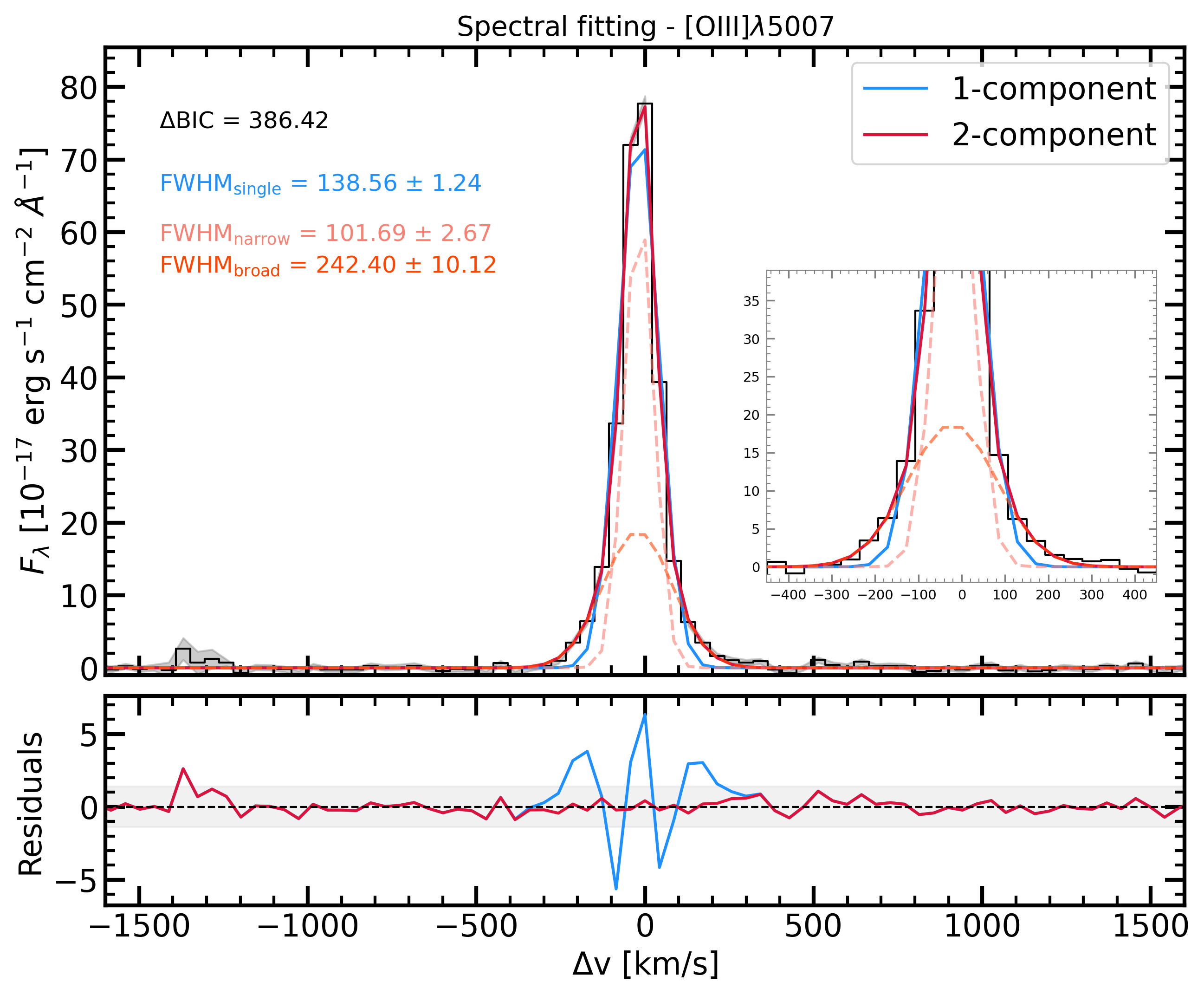}
        \end{subfigure}%
        \begin{subfigure}{0.5\textwidth}
        \centering
        \includegraphics[width=0.99\linewidth]{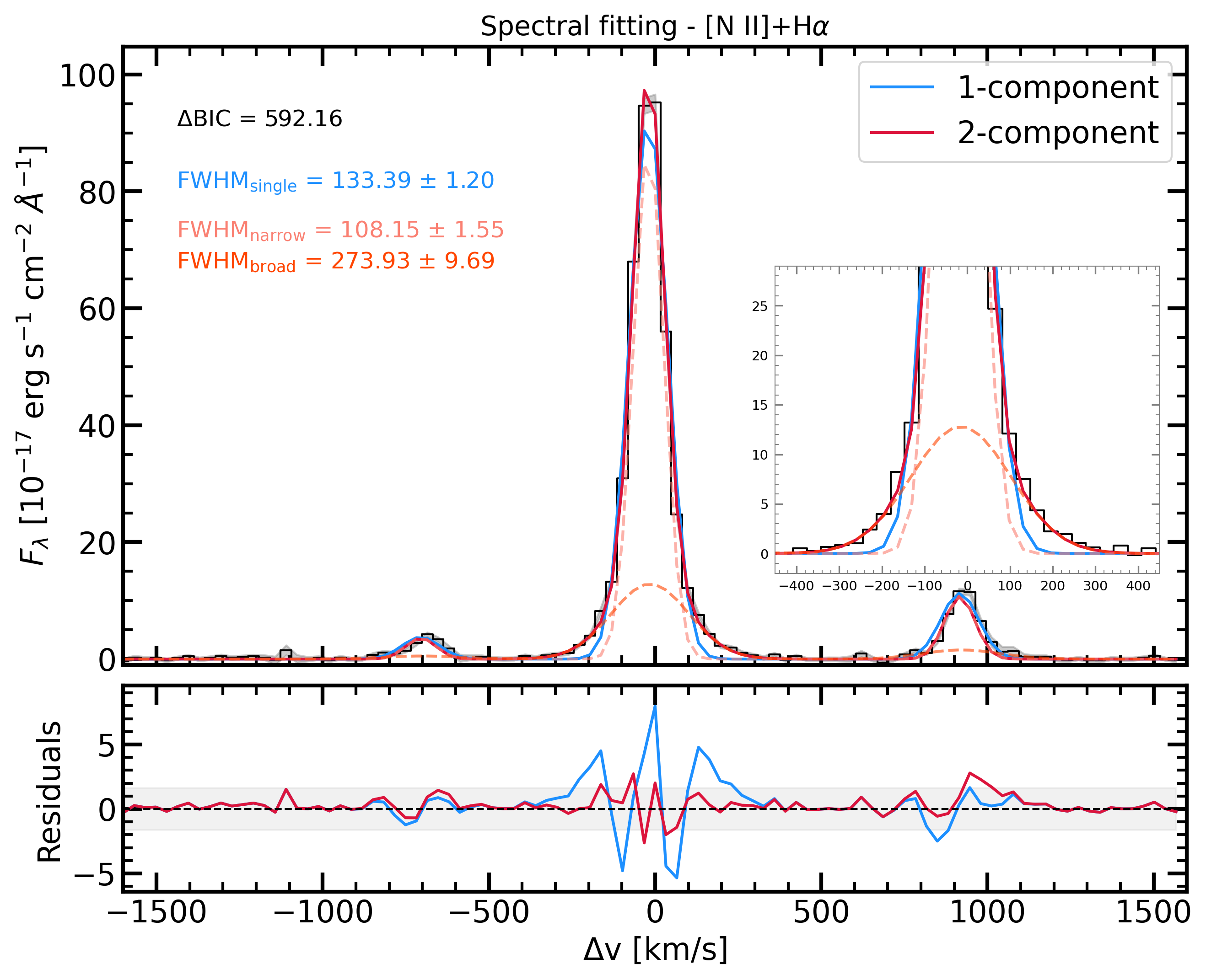}
        \end{subfigure}
        \caption{Example of spectral fitting for [O~III] (left) and [N~II]+H$\alpha$ (right) emission lines from the same target as in Fig.~\ref{fig:spectrum_example}. For each figure, the top panels show the continuum-subtracted spectra (black histogram) as a function of the velocity separation from the expected peak of the line. Single and double Gaussian fits on the lines are shown as blue and red solid lines, respectively. The salmon and orange dashed lines represent respectively the narrow and broad components from the double-Gaussian fit (no AGN component was preferred for the H$\alpha$ line). FWHM from all components (in km/s, uncorrected for instrumental broadening) and $\Delta$BIC are also reported. Finally, zoomed-in regions around the [O~III] and H$\alpha$ lines are shown as insets. The bottom panels display the corresponding residuals based on the single and double Gaussian fits. The gray band represents the residual rms in the case of single-component fit.}\label{fig:fit_example}
\end{figure*}

We fit each model with the \texttt{EMCEE} Markov Chain Monte Carlo (MCMC) approach \citep{Foreman13}, with a posterior probability of the form:
\begin{equation}
    \ln{P(\theta|\rm{y})} = \ln{\mathcal{L}(\rm{y}|\theta)} + \ln{\pi(\theta)},
\end{equation}
where $\theta$ represents the ensemble of model parameters, $\pi$ is the prior probability over the parameters, and $\mathcal{L}$ is the likelihood of the data given the model, as:
\begin{equation}
    \ln{\mathcal{L}} = -\frac{1}{2} \sum_i \left( \frac{F_{\lambda,i}(\theta) - y_i}{\sigma_{y,i}} \right)^2.
\end{equation}
For the priors, we imposed the velocity dispersions and integrated fluxes to be positive, and that $\sigma_{\rm{broad}} > \sigma_{\rm{narrow}}$ and $|\lambda_{\rm{narrow}}-\lambda_{\rm{broad}}|<20~{\AA}$ (corresponding to a velocity shift of $\sim1000~$km~s$^{-1}$), so to avoid nonphysical scenarios. We initialized the MCMC parameter space providing results from a least squares fit to the data using \texttt{scipy.optimize.curve$\_$fit} \citep{Virtanen20}, with initial parameter estimates based on the observed peak and line center. To ensure physical solutions, we constrained the narrow-component centroid to lie within $\pm5~\AA$ of the expected line center and imposed a minimum width of $1~\AA$. This provides a conservative initialization for the MCMC sampling, as this minimum width corresponds to $\sigma\sim60~\rm km~s^{-1}$ at [O~III]$\lambda5007$, larger than the typical DESI instrumental dispersion at the wavelengths probed by our sample ($\sigma_{\rm res}\sim30-40~\rm km~s^{-1}$; see Sect.~\ref{subsec:outflow_rate}). As stated above, the broad component was allowed a wider centroid range, within $\pm20~\AA$, to capture potential velocity shifts.

To choose among the two models, we used the Bayesian Information Criterion (BIC; \citealt{Liddle07}), defined as:
\begin{equation}
    \rm{BIC} = -2 \ln{\mathcal{L}_{\rm{max}}} + \textit{k} \ln{\textit{N}},
\end{equation}
where $\mathcal{L}_{\rm{max}}$ is the maximum likelihood from the posterior, $k$ represents the number of free parameters, and $N$ is the number of data points. We preferred the two-Gaussian model if:
\begin{equation}
    \Delta \rm{BIC} = \rm{BIC_{narrow}} - \rm{BIC_{narrow+broad}}\geq10.
\end{equation}
We additionally required the integrated flux of the broad component to be detected at $\mathrm{S/N}>2$, to exclude poorly constrained broad components from our outflow sample. We verified that adopting a more conservative $\mathrm{S/N}>3$ threshold does not affect any of the trends or conclusions presented below. Finally, from the posterior samples, we computed median values and uncertainties for each parameter based on the 16th, 50th, and 84th percentiles of the corresponding distributions.

For the H$\alpha$ line, we adopted a similar multi-component fitting approach, but also accounting for its proximity to the [N~II]$\lambda\lambda$6548,6583 doublet. In particular, we fit the [N~II]+H$\alpha$ complex by including two additional Gaussian for [N~II], fixing the wavelength offsets of the doublet relative to the H$\alpha$ (narrow or single component) centroid. To further reduce the number of free parameters, we estimated the amplitude of the [N~II]$\lambda$6583 line directly from the observed spectrum, by measuring the peak flux within a narrow window ($\pm5~\AA$) around the expected line center, and kept it fixed during the fitting. In addition, the [N~II] components were tied to share the same width as the narrow/single H$\alpha$ line, and the amplitude ratio of $\rm{[N~II]\lambda6583}/[N~II]\lambda6548$ was fixed to the theoretical value of 3 (e.g., \citealt{Acker89}). Furthermore, we introduced an additional very broad Gaussian component in Eq.~\ref{eq:gauss_model} to account for potential H$\alpha$ emission from a broad-line region (BLR) and prevent it from being attributed to the intermediate-width component associated with the outflow. This component was required to have a velocity dispersion at least 1.5 times as large as that of the broad H$\alpha$ component, in order to reduce degeneracies between the two components. The factor of 1.5 was adopted as a conservative fitting constraint rather than as a physically motivated threshold for BLR emission. We further required its centroid to lie within $\pm5~\text{\AA}$ of the narrow H$\alpha$ line, and its amplitude to be lower than that of the broad H$\alpha$ component to mitigate degeneracies. This extra Gaussian was not included for [O~III], as the forbidden nature of this transition precludes emission from the high-density BLR (e.g., \citealt{Husemann16,Singha22}). Accordingly, in our fitting procedure, [N~II]$\lambda\lambda$6548,6583 are modeled using narrow components only, tied in velocity and width to the narrow H$\alpha$ emission, with the broad and AGN components restricted to H$\alpha$ alone. We retained the additional very broad component only when statistically justified (i.e., when the model including it was preferred by $\Delta\mathrm{BIC}\geq10$). This was the case for 1159 sources, after applying additional quality cuts (see Sect.~\ref{subsec:cuts}). We stress that the presence of this additional component is not used as an independent AGN classification, but rather to account for possible BLR contamination when constraining the H$\alpha$ outflow component. 

We show in Fig.~\ref{fig:fit_example} the fitting results for the [O~III] and [N~II]+H$\alpha$ complex for the same galaxy shown in Fig.~\ref{fig:spectrum_example}. Additional examples of the spectral decomposition, spanning different stellar masses and BPT classifications, are presented in Appendix~\ref{app:spectral_fits}. These include an AGN-host galaxy for which the additional very broad H$\alpha$ component is statistically preferred, illustrating how our fitting procedure separates possible BLR emission from the intermediate-width component associated with the outflow.

\begin{figure*}[t]
    \begin{subfigure}{0.5\textwidth}
    \centering
        \includegraphics[width=0.972\linewidth]{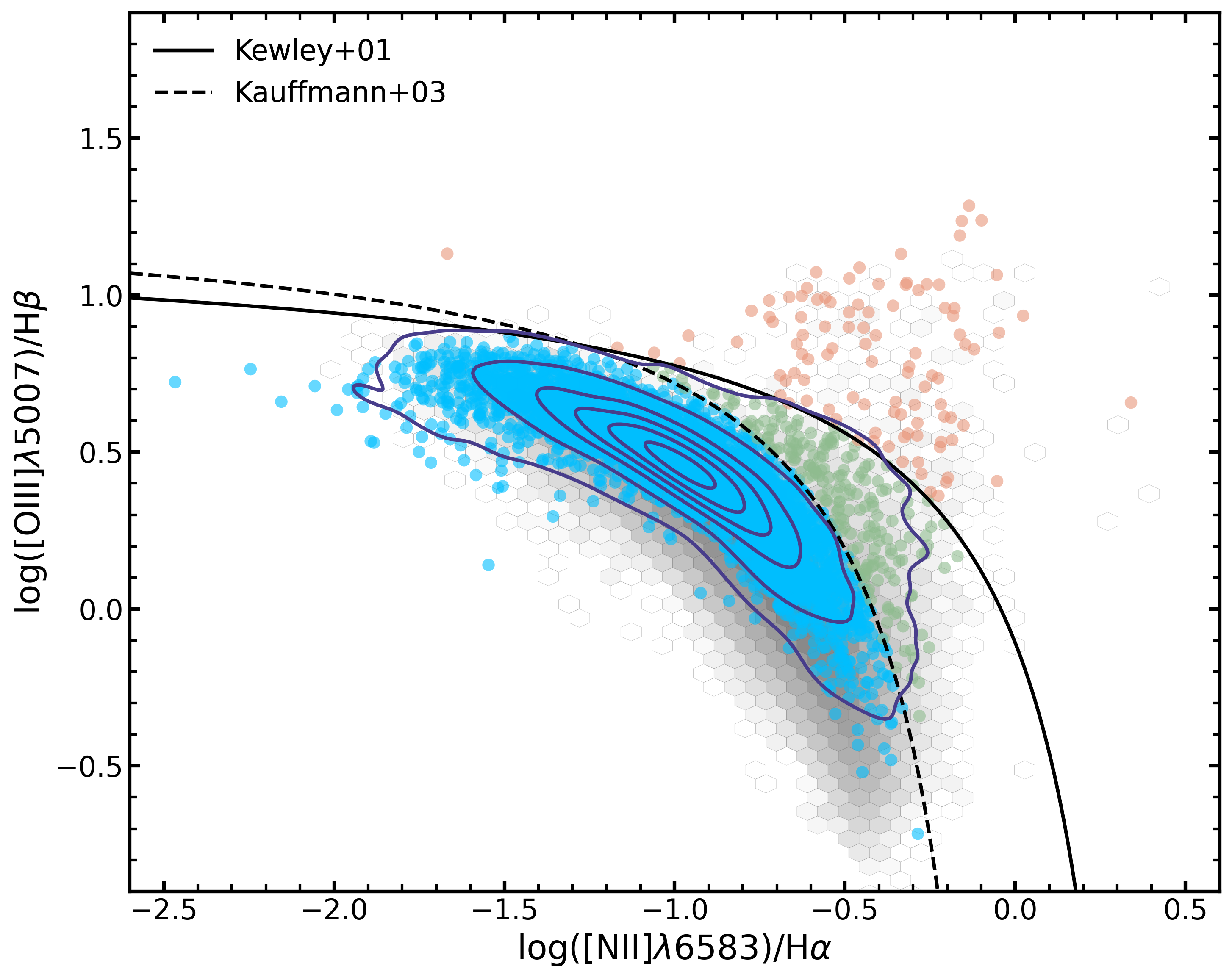}
        \end{subfigure}%
        \begin{subfigure}{0.5\textwidth}
        \centering
        \includegraphics[width=0.97\linewidth]{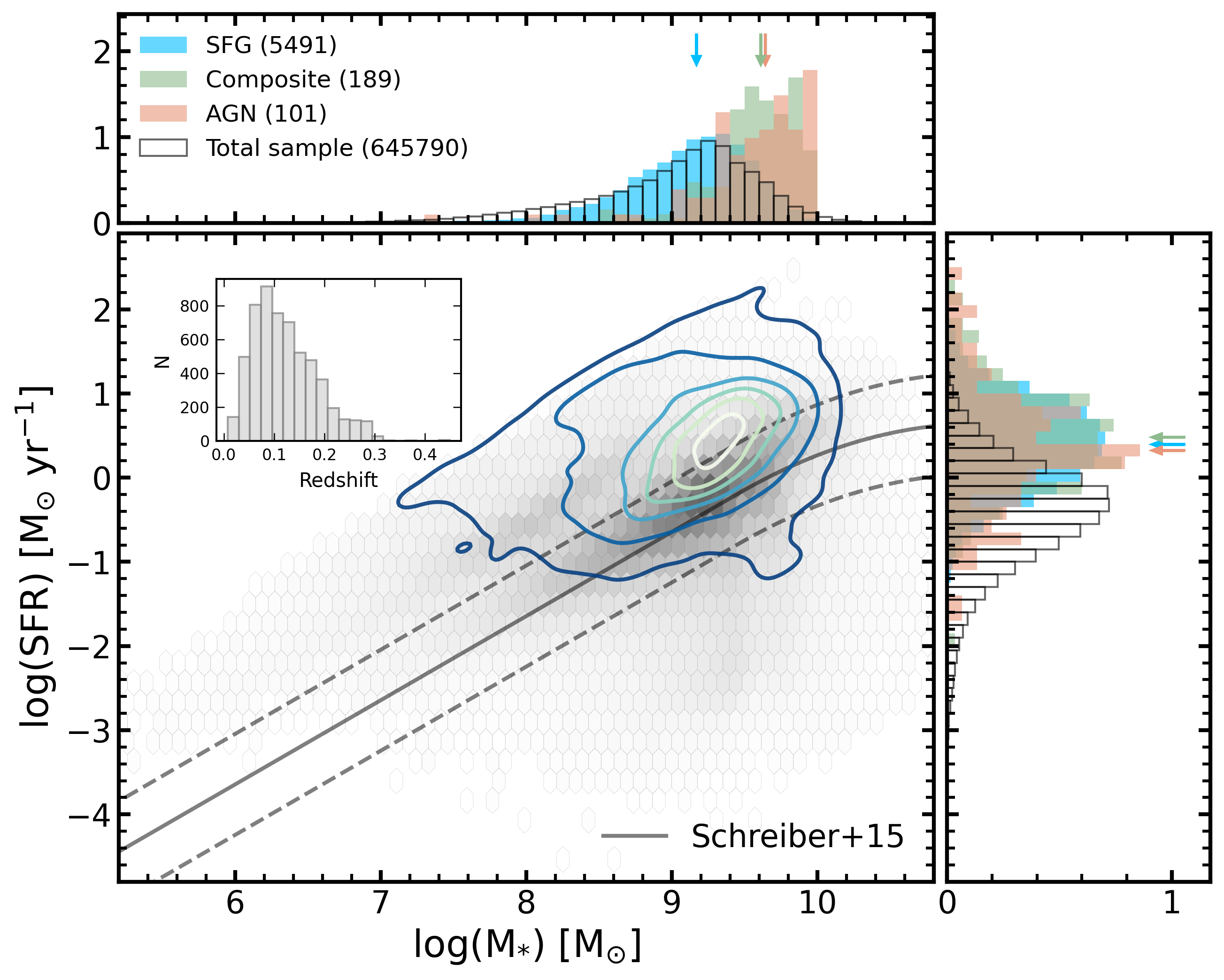}
        \end{subfigure}
        \caption{Physical and ionization properties of the dwarf galaxies with robust outflow detection. \textit{Left:} Classical BPT diagram showing [O~III]$\lambda5007$/H$\beta$ versus [N~II]$\lambda6583$/H$\alpha$. The black solid and dashed demarcations lines are from \cite{Kewley01} and \cite{Kauffmann03}, respectively, dividing the diagram into star-forming (blue), composite (green), and AGNs (red). Colored points and contours show the individual outflow candidates, while the grey hexagonal bins show the underlying dwarf parent population restricted to galaxies with $\mathrm{S/N}>3$ in all four emission lines entering the BPT diagnostic. All robust outflow candidates satisfy the same $\mathrm{S/N}$ criterion. The outermost contour encloses $1\%$ density level. Subsequent contours represent increasing fractions of the galaxy distribution, starting at $10\%$ and increasing in steps of $20\%$. \textit{Right:} distribution of the dwarf parent sample (grey hexagonal bins) and the individual outflow candidates (contours, at the same level as in the left panel) in the stellar mass-SFR plane. The solid and dashed grey lines correspond to the main-sequence from \cite{Schreiber15} at the median redshift of the dwarf parent sample ($z\sim0.1$), and its $\pm0.6~$dex offset, respectively. The inset shows the redshift distribution of the individual outflow candidates. The top and right histograms display the normalized distributions of outflow candidates by BPT classification (colored) compared to the general dwarf parent sample (open). The number of galaxies in each class and parent sample is also reported. Arrows indicate the median value of each BPT-classified distribution.}\label{fig:MS_BPT}
\end{figure*}

\subsubsection{Quality cuts}\label{subsec:cuts}
We further applied some quality cuts in order to ensure a robust outflow characterization. Following the quality-control approach of \cite{Pucha25}, we excluded sources with
\begin{equation}
    \sigma_{\rm{broad}} > 700~\rm{km~s^{-1}}.
\end{equation}
\cite{Pucha25} adopted the same threshold for the secondary [S~II] component to reject unreliable fitting solutions in noisy DESI spectra. Here, we apply it as a conservative upper limit to the broad outflow component, corresponding to $\mathrm{FWHM}_{\rm broad}\gtrsim1650~\rm{km~s^{-1}}$. These extreme widths populate only the tail of our fitted width distribution and are predominantly associated with low signal-to-noise spectra, for which the fitting procedure assigns a very broad Gaussian to noisy features at large velocity offsets, as confirmed by a posteriori visual inspection. Additionally, we required a detection for [O~II]$\lambda3727$, H$\beta$, and [N~II]$\lambda6584$ emission lines, that we use later for metallicity estimates (see Sect.~\ref{subsec:outflow_rate}). 

This selection yielded 6349 and 7696 outflow candidates when identified through the [O~III] and H$\alpha$ tracers, respectively. We note that the larger number of H$\alpha$-selected candidates may be partly due to the higher DESI spectral resolution at H$\alpha$ compared to [O~III], which can facilitate the identification of broad outflow components (e.g., \citealt{Hagedorn26}), as well as to the typically higher signal-to-noise ratio of the H$\alpha$ broad component. We subsequently cross-matched these two candidate sets to ensure a more robust outflow classification, resulting in a final sample of 5781 dwarf galaxies exhibiting clear evidence of ionized outflowing gas in both emission lines.

\subsubsection{Outflow properties}
Our conservative selection identifies robust outflow signatures in $\sim1\%$ of the dwarf parent sample. This fraction should not be interpreted as the intrinsic incidence of ionized outflows, since our selection criteria are designed to prioritize sample purity over completeness. Thanks to the large size of DESI, this nevertheless results in a statistically significant sample that enables us to investigate the properties of ionized outflows across a broad range of galaxy parameters.

To investigate the nature of outflows in dwarf galaxies, we first explored the dominant source of photoionization within the region sampled by the DESI fiber by looking at their distribution in the classic [N~II]/H$\alpha$ versus [O~III]/H$\beta$ BPT diagram \citep{Baldwin81}. Figure~\ref{fig:MS_BPT} (left panel) shows the position of our galaxy in the diagram along with separation lines between AGN and star-forming galaxies (SFGs) from \cite{Kewley01} and \cite{Kauffmann03}. Typically, galaxies below the \cite{Kauffmann03} line (dashed) are classified as SFGs, while those above the \cite{Kewley01} line (solid) need a non-stellar source of ionization, thus are classified as AGNs. Those in between the lines can show contributions from both AGN and star formation. The vast majority of our sample ($95\%$) lies in the star-forming region, with only small fraction of galaxies classified as composite ($3\%$) or AGNs ($2\%$). Nevertheless, due to the large sample size, even the relatively small AGN and composite populations contain sufficient galaxies for a statistical investigation of their outflow properties. These proportions are broadly consistent with those reported by \cite{Pucha25}. The somewhat larger fraction of AGN and composite systems in our sample may partly result from the different stellar-mass selection, as our sample extends to $\sim0.5$ dex higher stellar masses, where AGN activity becomes increasingly common (e.g., \citealt{Bongiorno12,Circosta18}). We caution that these classifications are based on the emission integrated within the DESI fiber aperture and therefore reflect the dominant ionization source within the region sampled by the fiber. In particular, star-formation emission may dilute the contribution from a weak nuclear AGN, potentially resulting in its classification as a star-forming galaxy based on the integrated line ratios. Similarly, the integrated BPT classification does not exclude a contribution from shock excitation to the broad emission, since the line ratios can be dominated by the brighter narrow component associated with H~II regions (e.g., \citealt{Allen08,Gallagher19}). 

We also note that the additional very broad H$\alpha$ component introduced to account for possible BLR contamination (Sect.~\ref{sec:method}) is statistically preferred in roughly one-third of the spectra of BPT-classified AGN/composite systems. While such a component may be consistent with nuclear activity, we do not use it as an independent AGN diagnostic, given the uncertainties associated with the spectral decomposition of H$\alpha$.

We also show in Fig.~\ref{fig:MS_BPT} (right panel) the sample distribution in redshift and in the stellar mass-SFR plane, along with the separate contribution from SFGs, composite, and AGNs. Composite systems and AGNs preferentially populate the high-mass end of our dwarf-galaxy sample, consistent with the well-known increase in the incidence of detectable AGN activity toward larger stellar masses (e.g., \citealt{Bongiorno12,Circosta18}). From Fig.~\ref{fig:MS_BPT} it is also evident that galaxies with outflow detections are preferentially located above the star-forming main sequence, in the starburst regime. This trend is consistent with the expectation that galaxies with elevated SFRs generate stronger outflows, and it is further reinforced by the fact that starburst galaxies exhibit brighter optical emission lines, making outflow signatures easier to detect in their spectra.

\section{Results}\label{sec:results}
\subsection{Outflow rates and efficiency}\label{subsec:outflow_rate}
To investigate the impact of ionized outflows in dwarf galaxies, we estimate here the outflow efficiency, or mass-loading factor ($\eta$), defined as the ratio between the gas mass-loss due to outflow (i.e., the outflow rate $\dot{M}_{\rm{out}}$) and star formation (i.e., $\eta\equiv\dot{M}_{\rm{out}}/\rm{SFR}$). The mass outflow rate can be defined as:
\begin{equation}\label{eq:outflow_mass}
    \dot{M}_{\rm{out}} = C~ \frac{M_{\rm{out}}~v_{\rm{out}}}{r_{\rm{out}}},
\end{equation}
where $C$ depends on the adopted outflow history, $M_{\rm{out}}$ is the mass of the outflowing gas, $v_{\rm{out}}$ is the outflow velocity, and $r_{\rm{out}}$ is its size (e.g., \citealt{Rupke05b,Lutz20}). For consistency with other works from the literature, we assume here $C=1$, implying a constant outflow rate over time (e.g., \citealt{Marasco23,Carniani24,Saldana25}). We note, however, that different geometries and outflow histories could provide outflow rates up to three times larger than those obtained with Eq.~\ref{eq:outflow_mass} (e.g., \citealt{Maiolino12,Cicone14}).


Since we do not have direct spatial information on the outflow extent, we assume $r_{\rm out}=1~\rm kpc$, motivated by spatially-resolved observations of ionized outflows in nearby star-forming and dwarf galaxies. Although the measured outflow extents show substantial object-to-object scatter, they are broadly distributed around characteristic scales of order $\sim1~\rm kpc$ (e.g., \citealt{Cresci17,Aravindan23,Marasco23}). We discuss the impact of this assumption and alternative prescriptions for $r_{\rm out}$ in Sect.~\ref{subsec:systematics}.

Following \cite{Rupke05b}, we compute the outflow velocities directly from the spectra, as:
\begin{equation}
    v_{\rm{out}} = |v_{\rm{broad}} - v_{\rm{narrow}}| + 2 \sigma_{\rm{broad,corr}},
\end{equation}
where $v_{\rm{broad}}$ and $v_{\rm{narrow}}$ are the velocity peaks of the broad and narrow components, respectively, and $\sigma_{\rm{broad,corr}}$ is the velocity dispersion of the broad component deconvolved by the instrumental resolution. DESI has a wavelength-dependent resolving power $R=\lambda/\Delta\lambda$, ranging from 2000 to 5200 from the bluest to the reddest wavelengths \citep{DESI22}. For each source, we estimated the instrumental FWHM at the observed wavelength of the considered emission line as $\mathrm{FWHM}{\rm_{res}}=c/R(\lambda{\rm_{obs}})$, and converted it into the corresponding Gaussian dispersion as $\sigma_{\rm res}=\mathrm{FWHM}_{\rm res}/2.355$. This results in typical values of $\sigma_{\rm res}\sim30-40~\mathrm{km~s^{-1}}$ at the observed wavelengths of [O~III] and H$\alpha$ for our sample. We then obtained the intrinsic broad-component dispersion as $\sigma^2_{\rm{broad,corr}}=\sigma^2_{\rm{broad}}-\sigma^2_{\rm{res}}$. The estimated outflow velocities are in the range $\mathrm{log(v_{out} / km~s^{-1})}\sim1.8-3.2$, similar to what found in local dwarf sources (e.g., \citealt{Marasco23}).

Following \cite{Carniani24}, we estimated the mass of the outflowing gas from both [O~III] and H$\alpha$, respectively as:
\begin{equation}\label{eq:mout_OIII}
    M_{\rm{out,[O~III]}} = 0.8\times10^8 \left(\frac{L^{\rm{broad}}_{\rm{[O~III]}}}{10^{44}~\rm{erg/s}} \right) {\left(\frac{Z_{\rm{out}}}{\rm{Z_{\odot}}} \right)}^{-1} \left(\frac{n_{\rm{out}}}{500~\rm{cm^{-3}}} \right)^{-1}~\rm{M_{\odot}},
\end{equation}
and
\begin{equation}
    M_{\rm{out,H\alpha}} = 0.8\times10^5 \left(\frac{L^{\rm{broad}}_{\rm{H\alpha}}}{10^{40}~\rm{erg/s}} \right) \left(\frac{n_{\rm{out}}}{100~\rm{cm^{-3}}} \right)^{-1}~\rm{M_{\odot}},
\end{equation}
with $Z_{\rm{out}}$ and $n_{\rm{out}}$ being the metallicity and electron density of the outflowing gas, and $L^{\rm{broad}}_{\rm{H\alpha}}$ and $L^{\rm{broad}}_{\rm{[O~III]}}$ representing the luminosities of the broad component of H$\alpha$ and [O~III] corrected for dust, respectively. Here, dust correction was applied by assuming a \cite{Cardelli89} extinction law with $R_{\rm{V}}=3.1$, and by computing the observed H$\alpha$/H$\beta$ flux ratio estimated from \texttt{FastSpecFit}, with an intrinsic ratio of 2.86 based on Case-B recombination \citep{Osterbrock06}\footnote{Although an intrinsic ratio of 3.1 is commonly adopted for AGN narrow-line emission (e.g., \citealt{Davies20}), we retain 2.86 for consistency across our predominantly star-forming sample. Nevertheless, we verified that adopting 3.1 for AGN-host galaxies does not significantly affect the main trends or conclusions of our analysis.}.

In principle, the electron density in the outflow can be estimated from diagnostic emission lines ratios, such as [S~II]$\lambda$6716/[S~II]$\lambda$6731. Following \cite{Sanders16}, we used the [S~II] lines to compute the global electron density in our galaxies, finding an average value of $\sim70~\rm{cm^{-3}}$, in agreement with results from local dwarf galaxies \citep{Marasco23}. However, as shown by \cite{Marasco23}, the electron density in the outflow component can be much larger than that in the galaxies disks. For this reason, and given that the outflow component was not detected in most of our galaxies for these diagnostic lines, we assume here $n_{\rm{out}}=380~\rm{cm}^{-3}$, as typically found for local and high-$z$ dwarf galaxies (e.g., \citealt{Marasco23,Saldana25}). Similarly, we assumed the outflow metallicity in Eq.~\ref{eq:mout_OIII} to be equal to the ISM metallicity \citep{HamelBravo24} that was estimated following \cite{Scholte24}. In brief, we derived dust-corrected R23 (i.e., ($\rm{[O~III]}\lambda\lambda4959,5007+\rm{[O~II]}\lambda3727$)/H$\beta$) and N2 (i.e., $\rm{[N~II]}\lambda6584$/H$\alpha$) line diagnostics, relating them to metallicities using calibrations from \cite{Nakajima22} and \cite{Denicolo02}, respectively. We then jointly minimized the residuals between the observed R23 and N2 diagnostics and their polynomial calibrations in order to derive the final metallicity for each galaxy, finding values in the range $7.3 < 12+\rm{log(O/H)} < 8.8$. We caution that the adopted metallicity calibrations were primarily developed for star-forming galaxies and may introduce additional systematic uncertainties for AGN-host and composite sources, where different ionization mechanisms can affect the observed emission-line ratios. To assess the impact of these uncertainties, we tested an alternative metallicity calibration specifically developed for AGN-ionized gas (e.g., \citealt{Carvalho20}) focusing on AGN-host galaxies, where the different ionization conditions may particularly affect the applicability of calibrations derived for star-forming gas. We find broadly consistent results between the two prescriptions and therefore retain our original metallicity estimates throughout the analysis, while acknowledging their limitations.

For the remainder of the paper, we adopt the outflow properties inferred from [O~III], while the corresponding H$\alpha$ estimates are used as an independent consistency check. The two tracers yield strongly correlated outflow properties and the same qualitative trends (see Appendix~\ref{app:oiii_ha_comparison}).

\begin{figure}[t!]
    \begin{center}
	\includegraphics[width=\columnwidth]{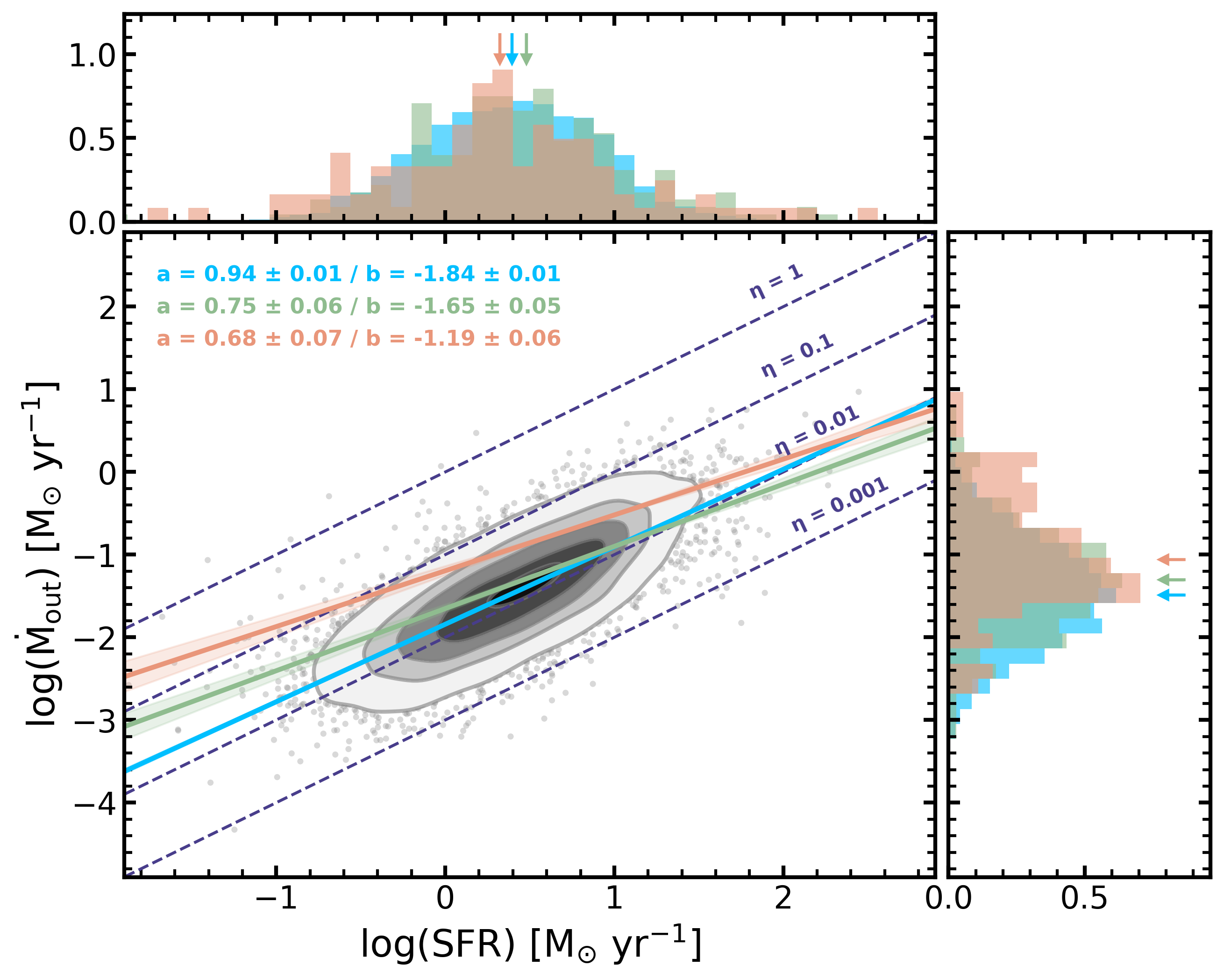}
	\end{center}
    \caption{Ionized outflow rate as a function of the SFR for the individual outflow candidates. Contours start at 10\% density levels, increasing in steps of 20\%. Top and right panels report the SFR and outflow rate distributions, respectively, for SFGs, composite systems, and AGNs, with arrows indicating their median values. The dashed purple lines display different values for the mass-loading factor. The solid lines with shaded areas show the results of a linear regression on the galaxies divided by their BPT classification, with slope and intercept (and corresponding uncertainties) reported on the top-left corner.}
    \label{fig:mout_sfr}
\end{figure}

We show in Fig.~\ref{fig:mout_sfr} the ionized outflow rate as a function of the SFR for the individual outflow candidates, as well as the parameter space covered by the outflows' mass loading factors. We found outflow efficiencies in the range $\sim0.001-1$, with a median value $\eta\sim0.01$, in agreement with findings from smaller samples of low-mass nearby and more distant sources (e.g., \citealt{Concas22,Marasco23,Hagedorn26}). This is $\sim2$ orders of magnitude smaller than what obtained from atomic and molecular outflows in local dwarfs and normal high-$z$ SFGs (e.g., \citealt{Ginolfi20,Herrera21,Romano23}), suggesting that the ionized phase may contribute only in a small percentage to the total outflow mass and energy budget (e.g., \citealt{Fluetsch21}). We caution that the absolute normalization of ionized mass-outflow rates, and hence of $\eta$, is subject to potentially large systematic uncertainties associated with assumptions on the physical conditions and spatial extent of the outflowing gas (e.g., \citealt{Harrison18,Rodriguez26}); we discuss the main uncertainties relevant to our analysis in Sect.~\ref{subsec:systematics}. Nonetheless, we still see a clear distinction in the outflow ionization rate and efficiency between SFGs, composite systems, and AGNs. To quantify that, we fitted the $\dot{M}_{\rm{out}}-\rm{SFR}$ relation with a linear regression of the form $\rm{log}(\dot{M}_{\rm{out}})=a\times\rm{log}(\rm{SFR})+b$ for each galaxy class, with $a$ and $b$ the slope and intercept of the relation, respectively. As reported in Fig.~\ref{fig:mout_sfr}, the slope and intercept get flatter and higher, respectively, going from SFGs to AGNs. This implies that AGNs drive stronger outflows with larger mass-loading factors and mass rates per fixed SFR as compared to SFGs, in agreement with expectation from more massive local galaxies (e.g., \citealt{Cicone14,Fluetsch19}). We verified that this enhancement remains essentially unchanged when adopting the alternative AGN metallicity calibration of \cite{Carvalho20}, indicating that our result is not significantly affected by the choice between these metallicity prescriptions. Furthermore, a potential overestimate of the SFRs in AGN-host galaxies (Sect.~\ref{subsec:catalog}) would lead to underestimated mass-loading factors and could further strengthen the observed enhancement relative to SFGs at fixed SFR.

\subsection{CGM and IGM enrichment}
To investigate whether the ionized outflows in our galaxies are able to escape their gravitational potential enriching the CGM (or possibly even the IGM), we compare their velocities with the escape velocities associated with the galactic disk and dark-matter halo. For consistency, we evaluate both gravitational potentials at the same characteristic radius, which we take to be our fiducial outflow radius, $r=r_{\rm out}=1~\rm kpc$. This scale is also comparable to the observed effective radii of our galaxies. Using the $r-$band measured half-light radii from the DESI Legacy Imaging Surveys \citep{Dey19} and converting them to physical units at the redshift of each source, we find a median $r_{\rm eff}\sim1.4~\rm kpc$, with a 16th--84th percentile range of $\sim0.8$--$2.7~\rm kpc$.

For the galactic disk, following e.g., \cite{Fluetsch19,Carniani24}, we assumed a Hernquist potential \citep{Hernquist90}, with density profile:
\begin{equation}
\rho(r) = \frac{M}{2\pi} \frac{a}{r} \frac{1}{(r+a)^3},
\end{equation}
where $M=M_{*}+M_{\rm{gas}}$, and $a=r_{\rm{eff}}~(1+\sqrt{2})^{-1}$, being $r_{\rm{eff}}$ the half-light (or effective) radius of the galaxy. In addition, given that no information on $M_{\rm{gas}}$ is available for our galaxies, we assumed $M_{\rm{gas}}/M_{*}\sim1$, as found on average for local galaxies at log$(M_{*}/\rm{M_{\odot}})\sim9$, the median stellar mass of our sample (e.g., \citealt{Catinella18,Hunt20}). 
The velocity required to escape the galaxy disk, evaluated at $r_{\rm out}$, is therefore
\begin{equation}
v_{\rm esc,disk}(r_{\rm out})
=
\sqrt{2|\Phi(r_{\rm out})|}
=
\sqrt{\frac{2G M}{r_{\rm out}+a}} .
\label{eq:vesc_disk}
\end{equation}

To compute the velocity needed from the gas to directly escape the dark matter halo, we assumed a Navarro-Frenk-White potential \citep{Navarro96} with the following density profile:
\begin{equation}
\rho(r) = \frac{\rho_{\rm{crit}}(z)~\delta_{\rm{c}}}{(r/r_{\rm{s}})(1+r/r_{\rm{s}})^2},
\end{equation}
where $\rho_{\rm{crit}}(z)$ is the critical density of the Universe at redshift $z$, $\delta_{\rm{c}}$ the characteristic overdensity of the halo, and $r_{\rm{s}}=r_{\rm{halo}}/c$ the characteristic radius, with $r_{\rm{halo}}$ the virial radius, and $c$ the concentration parameter. The latter was determined using the redshift-dependent $c$--$M_{\rm halo}$ relation of \cite{Dutton14}, being $M_{\rm{halo}}$ the halo mass obtained from the galaxy stellar mass through the stellar-to-halo mass relation by \cite{Behroozi10}. Finally, we computed $r_{\rm{halo}}$ following e.g., \cite{Lokas01,Huang17}, as:
\begin{equation}
r_{\rm{halo}} = \left[ \frac{3M_{\rm{halo}}}{4\pi~200~\rho_{\rm{crit}}(z)} \right]^{1/3}.
\end{equation}
From this, the halo escape velocity computed at the same
characteristic outflow radius is:
\begin{equation}
v_{\rm esc,halo}(r_{\rm out})
=
\sqrt{
\frac{2M_{\rm halo}G}
{r_{\rm out}
\left[\ln(1+c)-c/(1+c)\right]}
\ln\left(1+\frac{r_{\rm out}}{r_{\rm s}}\right)
}  .
\label{eq:vesc_halo}
\end{equation}

\begin{figure}[t!]
    \begin{center}
	\includegraphics[width=\columnwidth]{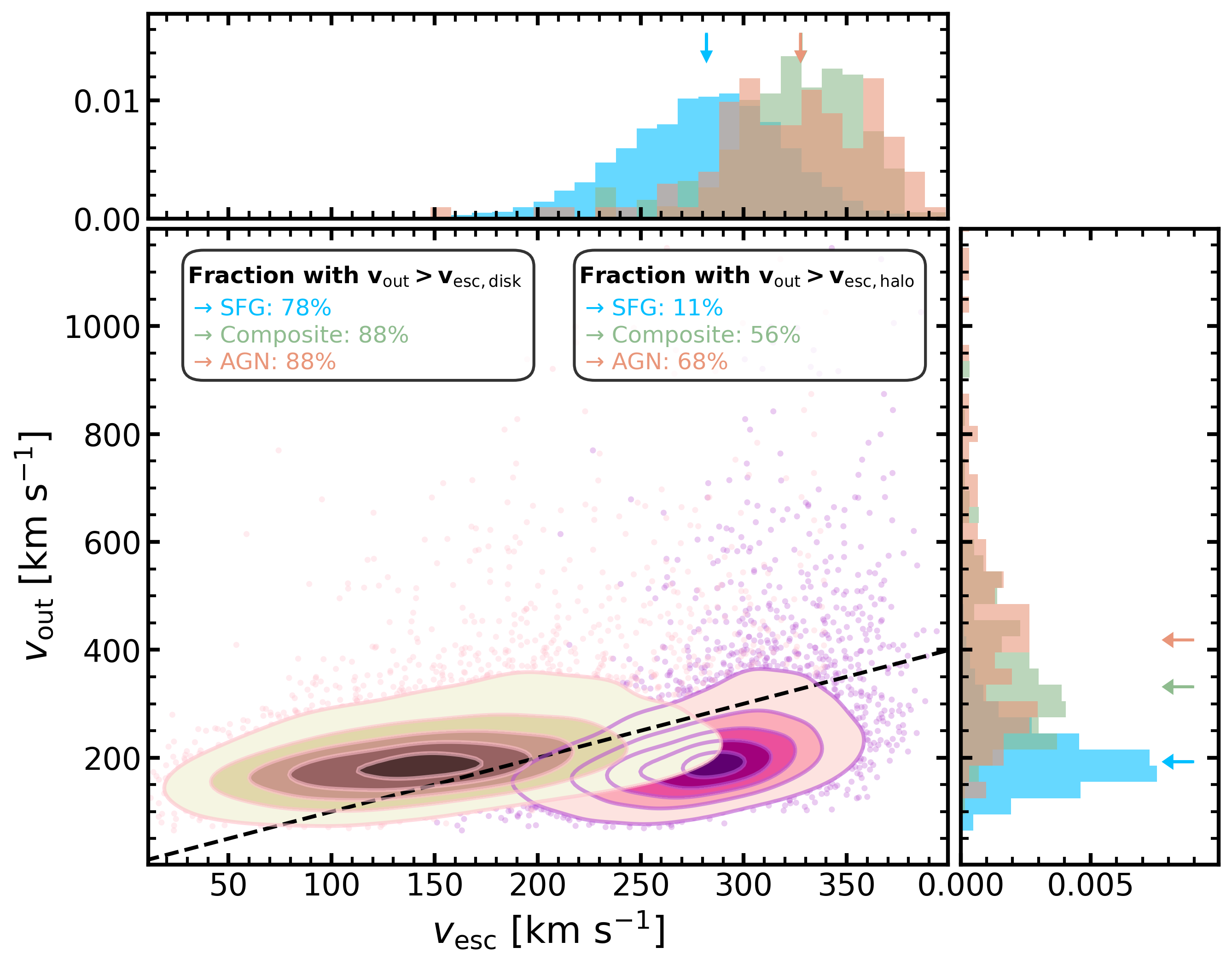}
	\end{center}
    \caption{Relation between outflow velocities and escape velocities for our individual outflow candidates. Pink and purple data points show the difference in the relations when computing the escape velocities from the galaxies' disks and dark matter halos, respectively. Contours start at $10\%$ density levels, increasing in steps of $20\%$. For both computations, we report the fraction of SFGs, composite, and AGNs with $v_{\rm{out}}>v_{\rm{esc}}$. In the case of $v_{\rm{esc,halo}}$, the corresponding velocity distributions (with arrows indicating their median values) are also shown in the top and right panels, to highlight the velocity separation for the different galaxies' classes (similar distributions, although shifted at smaller velocities, are also found for the relations with $v_{\rm{esc,disk}}$). The dashed black line corresponds to the $1:1$ relation.}
    \label{fig:vesc_vout}
\end{figure}

Figure~\ref{fig:vesc_vout} shows the comparison between the outflow velocities and the escape velocities associated with the galactic disk and dark-matter halo, derived from Eqs.~\ref{eq:vesc_disk} and \ref{eq:vesc_halo}, respectively. Both quantities increase with stellar mass, producing the broad correlation visible in the figure. A clear separation between SFGs, composite systems, and AGNs is also apparent, with the latter preferentially occupying the higher-mass end of the sample and showing a pronounced tail toward high outflow velocities ($v_{\rm out}>500~\rm km~s^{-1}$).

A large fraction of our galaxies have outflow velocities comparable to or larger than $v_{\rm esc,disk}$, with the fraction satisfying $v_{\rm out}>v_{\rm esc,disk}$ increasing from $\sim78\%$ for SFGs to $\sim88\%$ for both composite systems and AGNs. This suggests that the high-velocity ionized component can frequently escape the central galaxy and reach the CGM. The fractions decrease substantially when considering the dark-matter halo potential: only $\sim11\%$ of SFGs have $v_{\rm out}>v_{\rm esc,halo}$, compared with $\sim56\%$ and $\sim68\%$ of composite systems and AGNs, respectively. Thus, while ionized outflows can efficiently transport material from the galaxy into the CGM, most of the ionized gas associated with star-formation-driven outflows is expected to remain gravitationally bound to the halo and may eventually be recycled onto the galaxy. In contrast, the high-velocity ionized component in a substantially larger fraction of composite and AGN systems reaches velocities sufficient to escape the halo and potentially enrich the IGM.

We caution, however, that our adopted $v_{\rm out}$ represents a characteristic maximum velocity of the broad emission-line component, rather than the bulk velocity of the outflowing gas. The fractions with $v_{\rm out}>v_{\rm esc}$ should therefore not be interpreted as the fraction of the total ionized outflowing material that escapes the galaxy or halo, but rather as the fraction of systems in which the observed high-velocity ionized component reaches velocities sufficient to do so.

\section{Discussion}\label{sec:discussion}
\begin{figure*}[t!]
    \begin{center}
	\includegraphics[width=0.9\textwidth]{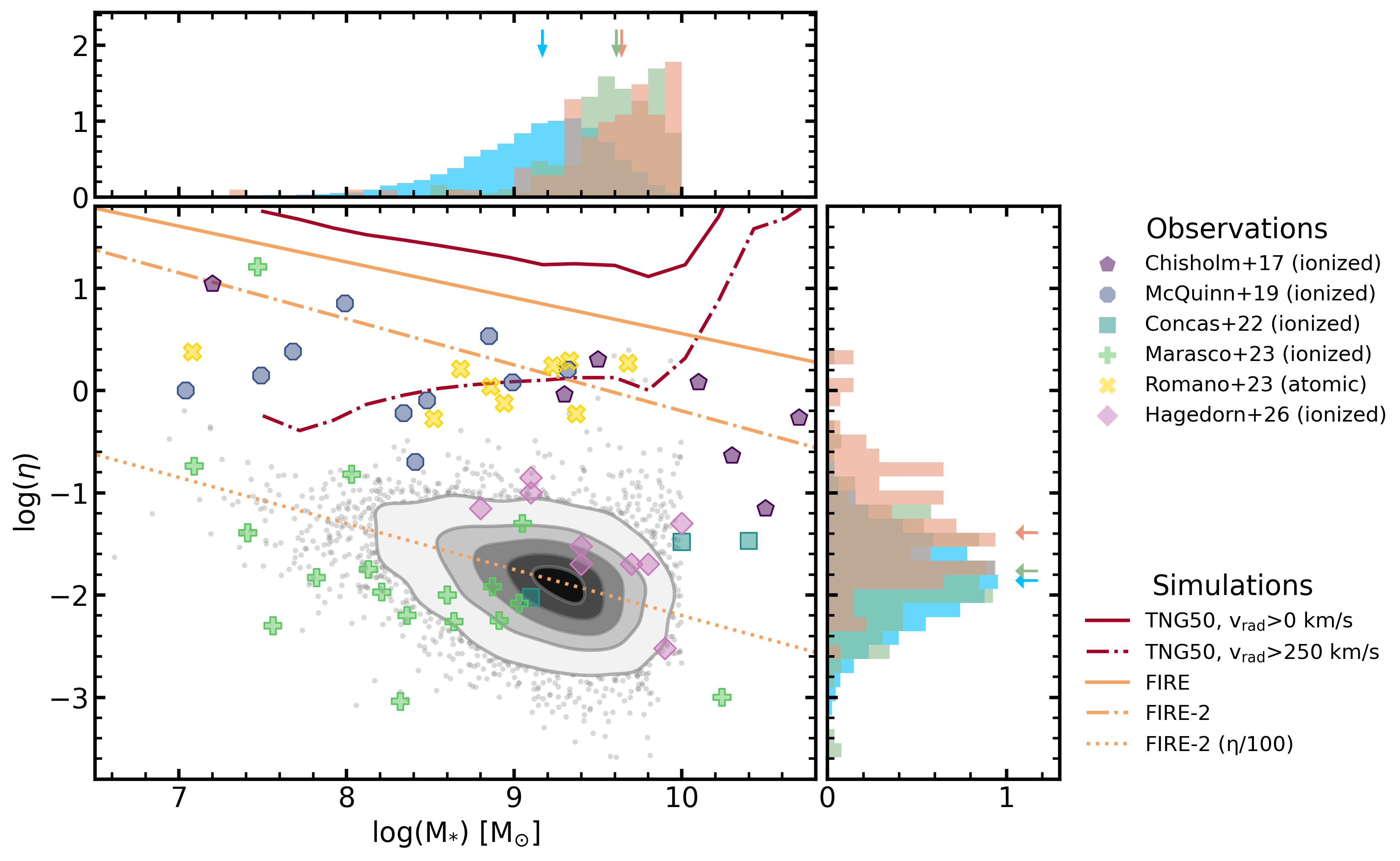}
	\end{center}
    \caption{Mass-loading factor as a function of stellar mass for our dwarf galaxies (grey points and contours). Contours start at 10\% density levels, increasing in steps of 20\%. Top and right panels report the parameters' distributions for SFGs, composite systems, and AGNs, with arrows indicating their median values. Coloured markers show the results of observational studies from \citet[pentagons]{Chisholm17}, \citet[octagons]{McQuinn19}, \citet[squares]{Concas22}, \citet[plus symbols]{Marasco23}, \citet[X symbols]{Romano23}, and \citet[diamonds]{Hagedorn26}, with the probed gas phase reported in parenthesis. Solid and dot-dashed lines show predictions from the Illustris TNG50 (\citealt{Nelson19}; brown solid and dashed-dot for $v_{\rm{rad}}>0~$km~s$^{-1}$ and $v_{\rm{rad}}>250~$km~s$^{-1}$, respectively), the FIRE (\citealt{Muratov15}; orange solid), and the FIRE-2 (\citealt{Pandya21}; orange dashed-dot) cosmological simulations. The orange dotted line shows the FIRE-2 simulation rescaled by a factor 100, for reference.}
    \label{fig:eta_mstar}
\end{figure*}

\subsection{Comparison with previous studies}
We show in Fig.~\ref{fig:eta_mstar} the variation of $\eta$ with the stellar mass, and compare our measurements with previous observational and theoretical studies covering a similar parameter space. This figure highlights a wide scatter in the observations, likely driven by differences in methodology and the probed outflow phases. Nonetheless, our results are broadly consistent with recent studies of ionized outflows in both local and $z\sim2$ dwarf galaxies (e.g., \citealt{Concas22,Marasco23,Hagedorn26}). In particular, \cite{Marasco23} made use of MUSE data to model H$\alpha$ emission of 19 nearby starburst galaxies, finding mass-loading factors in the range $0.001-10$, in line with our measurements. Similar results were obtained by \cite{Concas22}, who employed physically-motivated models to separate the contribution of ionized outflows from the galaxies' rotating discs in stacked $z\sim2$ low-mass galaxies from the KLEVER survey \citep{Curti20,Hayden22}. At $\rm{log(M_{*}/M_{\odot})\sim9}$, they report an average $\eta\sim0.01$, again in excellent agreement with our findings. More recently, \cite{Hagedorn26} used medium-high resolution optical spectroscopy of local star-forming galaxies with $\rm{log}(M_{}/M_{\odot})=8.5-10$ to identify and characterize ionized outflows through H$\alpha$. They find a mean mass-loading factor of $\eta\sim0.07$, well within the range covered by our sample, providing an independent comparison at similar stellar masses.

On the other hand, other studies find significantly higher outflow efficiencies. For example, \cite{McQuinn19} characterized the strength of ionized outflows in 12 nearby dwarf galaxies using deep narrow-band H$\alpha$ imaging. They concluded that the majority of the star formation-driven material expelled from the galaxies will not escape into the IGM, but will likely remain bound to the galaxies' halo and re-accreted onto the galactic discs, in agreement with our findings. At the same time, they derived mass-loading factors from 0.2 to 7, which are $10-1000$ larger than what estimated in our work. Such discrepancy may be partially explained by the different methods involved: while \cite{McQuinn19} based the mass-loading factor computation on ionized gas morphology, we instead rely on its kinematical analysis (see more detailed discussion in \citealt{Marasco23}). Different gas tracers may also lead to different results. \cite{Chisholm17} used UV absorption line measurements from the Cosmic Origins Spectrograph to characterize galactic outflows in a small sample of 8 nearby galaxies, finding $\eta\sim1-20$ for sources with $\rm{log(M_{*}/M_{\odot})<10}$, more than a factor 100 larger than our median value. Likewise, in \cite{Romano23} we used [CII] emission to trace atomic outflows in local dwarf galaxies, finding a typical mass-loading factor $\eta\sim1$, almost two orders of magnitudes higher than what found in our work, possibly pointing to a subdominant contribution of ionized gas to the overall outflow budget (e.g., \citealt{Fluetsch21}). We note however that, as discussed in Sect.~\ref{subsec:systematics}, the absolute normalization of observationally inferred mass-loading factors is subject to substantial systematic uncertainties, arising from assumptions about the outflow geometry and extent, the physical conditions of the gas, and the conversion of the adopted emission-line tracer into outflowing gas mass. Differences in these assumptions, in addition to the adopted methodology and gas tracer, can therefore contribute significantly to the large scatter among literature measurements. 

\begin{figure*}[t!]
    \begin{center}
	\includegraphics[width=0.9\textwidth]{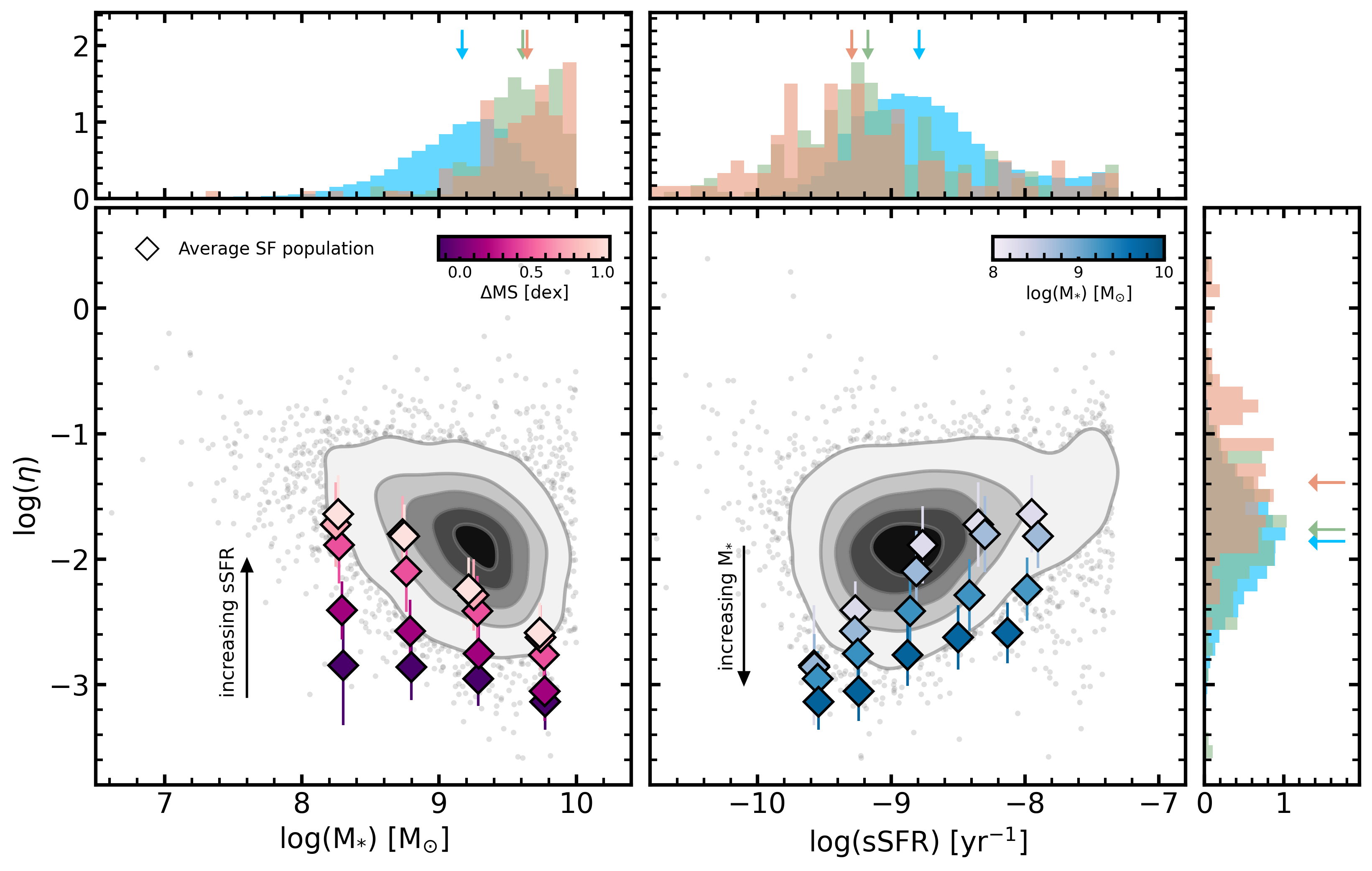}
	\end{center}
    \caption{Mass-loading factor as a function of stellar mass (left) and specific star-formation rate (right) for our sample (grey points and contours). Contours start at 10\% density levels, increasing in steps of 20\%. Top and right panels show the normalized distributions of stellar mass, sSFR, and mass-loading factor for star-forming (blue), composite (green), and AGN (orange) systems, with arrows indicating the corresponding median values. Large diamonds show average mass-loading factors derived from stacked DESI spectra of the star-forming dwarf population following Romano \& Saintonge (2026, subm.), measured in bins of stellar mass and offset from the star-forming main sequence ($\Delta$MS). In the left panel, the stacked measurements are colour-coded by $\Delta$MS, while in the right panel they are colour-coded by stellar mass. Error bars indicate uncertainties on $\eta$. In the left panel, stacked galaxies exhibit systematically lower mass-loading factors than individually detected outflows at fixed stellar mass, although they progressively approach the individual detections toward larger $\Delta$MS. In the right panel, both individual and stacked measurements show increasing outflow efficiency with increasing sSFR, suggesting that elevated star-formation activity and burstiness enhance feedback efficiency and may drive the transition toward more efficient galactic winds. Black arrows indicate the direction of increasing sSFR (left) and stellar mass (right).}
    \label{fig:eta_mstar_ssfr}
\end{figure*}

\subsection{Comparison with simulations and driving mechanisms}
Figure~\ref{fig:eta_mstar} shows that the outflow mass-loading factor decreases with increasing stellar mass. This trend is also predicted by theoretical models, with the slope of the relation providing insights into the physical driving mechanisms of outflows in star-forming galaxies, i.e., whether they are predominantly ‘energy'- or ‘momentum'-driven. In momentum-driven scenarios, stellar feedback rapidly loses thermal energy through radiative cooling, and the outflow is primarily accelerated by momentum injection from supernovae and radiation pressure on dust grains associated with young stellar populations. This leads to relatively weak outflows with scaling relations of $\eta\propto v_{\rm{out}}^{-1}$ and $\eta\propto M_{*}^{-1/3}$. Conversely, if supernovae occur in clustered star-forming regions, the injected energy can remain efficiently coupled to the ISM before significant cooling occurs, generating more powerful energy-driven winds with steeper scalings of $\eta\propto v_{\rm{out}}^{-2}$ and $\eta\propto M_{*}^{-2/3}$ (e.g., \citealt{Chevalier85,Murray05,Oppenheimer06,Murray11,Mitra15,Muratov15}). As a result, observational measurements of mass-loading factors can provide important constraints on the dominant feedback mode operating in galaxies.

In Fig.~\ref{fig:eta_mstar} we compare our results with predictions from the Illustris TNG50 \citep{Nelson19,Pillepich19} and the Feedback in Realistic Environments (FIRE; \citealt{Hopkins14,Muratov15,Pandya21}) cosmological simulations. At fixed stellar mass, these models predict differences in outflow efficiency up to two orders of magnitude, likely reflecting both the adopted feedback prescriptions and the numerical characteristics of the simulations. 

In the case of TNG50, outflow mass-loading factors were computed by \cite{Nelson19} at a fixed distance of 10~kpc from the galaxy, and with energy-driven winds as input. We show their results in Fig.~\ref{fig:eta_mstar} in the case of radial velocities $v_{\rm{rad}}>0$~km~s$^{-1}$ (solid brown line) and $v_{\rm{rad}}>250$~km~s$^{-1}$ (dashed-dot brown line), representing upper and lower bounds, respectively. As shown, the $v_{\rm{rad}}>0$~km~s$^{-1}$ case produces the highest mass-loading factors across the whole mass range, following a trend broadly similar to our data but systematically more than two orders of magnitude higher, in line with results by e.g., \cite{Concas22,Marasco23}. The more conservative $v_{\rm{rad}}>250$~km~s$^{-1}$ case better reproduces some literature measurements and approaches the upper envelope of our observations, but predicts a qualitatively different behaviour, with mass-loading factors becoming increasingly efficient toward higher stellar masses, in contrast with the observed decrease of $\eta$ with $M_{*}$. Similarly, we show the predictions by \cite{Muratov15} and \cite{Pandya21} for the FIRE and FIRE-2 simulations, respectively. As highlighted by \cite{Pandya21}, the differences between these two simulations primarily arise from how the mass outflow rate is computed, rather than from changes in feedback prescriptions (see also discussion in \citealt{Kado24}). Specifically, \cite{Muratov15} computed $\eta$ at a distance of $0.2-0.3~R_{\rm{vir}}$ (with $R_{\rm{vir}}$ the virial radius), and assuming $v_{\rm{rad}}>0$~km~s$^{-1}$, while \cite{Pandya21} used a different velocity criterion at $0.1-0.2~R_{\rm{vir}}$ to select only gas in the outflow and avoid slowly moving particles in turbulent motions. Since this slow component is included in \cite{Muratov15}, their predictions on mass-loading factors can be considered as upper limits from FIRE simulation. Interestingly, our observed dependence of mass-loading factor on stellar mass appears comparable to, or even steeper than, the FIRE-2 relation, whose scaling is broadly consistent with energy-driven winds (\citealt{Pandya21}; see discussion in Sect.~\ref{sec:stacking}). However, reproducing the normalization of the bulk of our measurements requires reducing the FIRE-2 relation by approximately a factor $\sim100$ (Fig.~\ref{fig:eta_mstar}). 

We note that discrepancies between our results and cosmological simulations could be in principle alleviated if considering systematic uncertainties in the observational estimates of $\eta$. For instance, higher outflow velocities, smaller gas densities, or more compact outflow sizes would all act to increase $\eta$, bringing it closer to predictions. However, such assumptions are unlikely and would be inconsistent with current observations of nearby and high-$z$ dwarf galaxies (e.g., \citealt{Romano23,Marasco23,Carniani24,Saldana25}). The assumed outflow geometry could also contribute, with spherical or bi-conical geometries yielding up to a factor of $\sim3$ larger outflow rates compared to our estimates (e.g., \citealt{Cicone14,Lutz20}), but this is still insufficient to explain the large gap with simulations. A more likely explanation lies in the multi-phase nature of outflows. As already suggested for local ULIRGs, starburst galaxies, AGN hosts, and high-$z$ galaxies (e.g., \citealt{Cicone18,Fluetsch19,Herrera19,Fluetsch21,Hernandez26}), the ionized gas phase we are probing in this work likely represents only a minor fraction of the total mass and energy budget of outflows, even in the low-mass regime. If this is the case, then our low $\eta$ values simply reflect the reduced role of ionized gas compared to the atomic and molecular phases. This interpretation would also explain why some simulations in Fig.~\ref{fig:eta_mstar} can partially reproduce observations for atomic outflowing gas in dwarf galaxies.


\subsection{Comparison with the average star-forming population}\label{sec:stacking}
The analysis presented so far is based on individually detected outflows and therefore preferentially selects galaxies in which the broad component is sufficiently prominent to be identified in individual DESI spectra. To place these detections in the context of the average star-forming dwarf population,
we additionally performed a spectral stacking analysis. This allows us to recover weaker outflow signatures that would otherwise remain undetected in individual spectra and to investigate how the average outflow efficiency varies with stellar mass and star-formation activity.


We follow the stacking methodology described in Romano \& Saintonge (subm.). Galaxies are divided into bins of stellar mass ($0.5$ dex) and offset from the star-forming main sequence ($\Delta{\rm MS}$; $0.3$ dex). In short, within each bin, the DESI spectra are corrected for Galactic extinction, shifted to the rest frame, resampled onto a common wavelength grid, normalized, and mean-stacked. The stellar continuum of each stack is then modelled and subtracted using FastSpecFit. While Romano \& Saintonge (subm.) use H$\alpha$ to characterize the stacked outflows, here we perform the analysis on the [O\,III]$\lambda5007$ emission line in order to ensure a direct and homogeneous comparison with the individually detected [O\,III] outflows analysed throughout this work. We fit the stacked [O\,III] profiles with both single- and two-component Gaussian models and adopt the latter when the inclusion of a broad component is statistically favoured ($\Delta{\rm BIC}\geq10$). For stacks with a detected broad component, we derive the outflow properties following the same prescriptions adopted for the individual outflow candidates, including the same assumptions for the outflow radius, electron density, and metallicity (see Sects.~\ref{subsec:fitting} and \ref{subsec:outflow_rate}). This ensures that differences between the individual and stacked measurements are not driven by different assumptions in the conversion from [O\,III] emission to outflow properties.

The resulting average mass-loading factors are shown as large diamonds in Fig.~\ref{fig:eta_mstar_ssfr}, colour-coded by $\Delta{\rm MS}$ in the left panel and by stellar mass in the right panel. The stacked measurements generally lie below the individually detected outflows. At fixed stellar mass, the average star-forming population therefore exhibits lower ionized mass-loading factors than galaxies in which an outflow can be identified directly. This difference is expected, since the individual selection preferentially identifies systems with stronger broad components, whereas stacking probes the average galaxy population, including galaxies with substantially weaker outflow signatures. Importantly, the discrepancy progressively decreases toward larger $\Delta{\rm MS}$, with stacks of galaxies further above the main sequence approaching the region occupied by the individual detections. This is consistent with our individual outflow sample being preferentially composed of galaxies with elevated star-formation activity (Fig.~\ref{fig:MS_BPT}).


Romano \& Saintonge (subm.) also found that the slope of the $\eta-M_\star$ relation steepens systematically with increasing $\Delta$MS, evolving from values consistent with momentum-driven outflows in main-sequence galaxies toward slopes approaching the energy-driven expectation in starburst systems. In that framework, enhanced star-formation activity promotes clustered supernova feedback and more efficient coupling between stellar feedback and the ISM, producing increasingly efficient outflows in low-mass galaxies. The relation derived from the individually detected outflows in this work appears broadly consistent with this picture: although our sample includes galaxies spanning a range of $\Delta$MS values, the measured trend lies closer to the energy-driven regime than to the momentum-driven expectation. This may naturally arise because individually detected broad components preferentially select galaxies experiencing stronger episodes of feedback, while the inclusion of lower-$\Delta$MS systems likely produces a somewhat shallower overall slope compared to the most extreme stacked starburst bins.

This interpretation is further supported by the relation between mass-loading factor and specific star-formation rate (sSFR) shown in Fig.~\ref{fig:eta_mstar_ssfr}. While the dependence on stellar mass remains visible, a positive trend emerges between $\eta$ and sSFR, with increasingly efficient outflows observed in galaxies with larger SFRs relative to their stellar mass. The stacked measurements follow the same behaviour and move toward the region occupied by individual detections as $\Delta$MS increases. A positive correlation between mass-loading factor and sSFR is also predicted by recent numerical simulations. For example, \cite{Martin26} find that higher sSFR values correspond to larger outflow efficiencies and stronger burst-driven feedback episodes, with the most efficient winds occurring in galaxies exhibiting highly variable star-formation histories. In this framework, elevated sSFR is interpreted as a manifestation of increased star-formation burstiness, which enhances feedback coupling and leads to stronger mass ejection.

Taken together, these results suggest that individually detected ionized outflows do not necessarily represent typical feedback conditions in low-mass galaxies. Instead, they preferentially trace the high-efficiency tail of the outflow population associated with phases of enhanced star formation. Conversely, stacked spectra recover lower, more representative average outflow efficiencies across the broader star-forming population. The decreasing difference between the stacked and individual measurements toward large $\Delta{\rm MS}$ further suggests that this contrast is primarily related to galaxy selection and star-formation activity rather than to differences in the methodology used to derive the outflow properties. Moreover, the behaviour of the individually detected outflows is consistent with the picture proposed by Romano \& Saintonge (subm.), in which increasingly bursty star formation is associated with a transition from outflow scalings characteristic of momentum-driven winds toward those expected for more efficiently energy-driven feedback.

\subsection{Systematic uncertainties on the mass-loading factors}\label{subsec:systematics}
The derivation of ionized mass-loading factors from spatially-unresolved spectroscopy necessarily relies on assumptions about the physical conditions and spatial extent of the outflowing gas. These assumptions can introduce substantial systematic uncertainties in the absolute normalization of the inferred outflow rates and energetics (e.g., \citealt{Harrison18}). Here, we focus on two uncertainties particularly relevant to our analysis: the assumed outflow radius and the conversion of [O~III] luminosity into ionized gas mass. Importantly, while these assumptions can affect the absolute values of $\eta$, our main conclusions rely primarily on relative trends with galaxy properties.\\

\noindent
\textbf{Outflow radius.} One of the main uncertainties in estimating mass-outflow rates from spatially unresolved spectroscopy is the characteristic outflow radius, since $\dot{M}_{\rm out}\propto r_{\rm out}^{-1}$. Our fiducial choice of $r_{\rm out}=1~\rm kpc$ is motivated by spatially-resolved observations of ionized outflows in nearby star-forming and dwarf galaxies. Although the measured extents exhibit substantial object-to-object scatter and depend on the adopted definition of outflow size, they are broadly distributed around scales of order $\sim1~\rm kpc$ (e.g., \citealt{Cresci17,Aravindan23,Marasco23}). To assess the impact of adopting a fixed radius, we repeated our analysis using, for each galaxy, the physical radius subtended by the DESI fiber as an alternative characteristic scale, following a similar aperture-based approach to that adopted in other spatially-unresolved studies (e.g., \citealt{Liu20,Rodriguez26}). We stress that the fiber radius does not represent a direct measurement of the physical outflow extent and does not account for its intrinsic geometry or spatial distribution. Nevertheless, the resulting mass-loading factors remain tightly correlated with those obtained assuming $r_{\rm out}=1~\rm kpc$, and the main trends presented in this work remain unchanged. This is partly because, around the median redshift of our sample ($z\sim0.1$), the physical DESI fiber radius is itself of order $1~\rm kpc$. We therefore conclude that, while the assumed radius contributes directly to the systematic uncertainty in the absolute normalization of $\eta$, adopting an aperture-dependent radius does not drive the trends investigated here.\\

\noindent
\textbf{[O~III]-based ionized gas masses.} A further uncertainty arises from our use of [O~III] to estimate the ionized outflow mass. Unlike hydrogen recombination lines, the conversion between [O~III] luminosity and ionized gas mass depends on the oxygen abundance and implicitly on the fraction of oxygen in the O$^{++}$ ionization state (e.g., \citealt{CanoDiaz12,Carniani15,Harrison18,Kakkad20,Rodriguez26}). Variations in ionization conditions with galaxy properties could therefore affect the inferred mass-loading factors. We can test the importance of this effect using H$\alpha$, which does not require an ionization correction. As shown in Appendix~\ref{app:oiii_ha_comparison}, the outflow properties independently inferred from H$\alpha$ and [O~III] are strongly correlated, and adopting the H$\alpha$-based estimates does not alter the main trends presented in this work. We therefore conclude that the use of [O~III] is unlikely to drive our results, although it remains an additional source of systematic uncertainty in the absolute normalization of the inferred ionized gas masses.

\section{Summary and conclusions}\label{sec:summary}
We investigated the properties of ionized outflows in a large statistical sample of low-mass ($\mathrm{log(M_{*}/M_{\odot}})\lesssim10$) galaxies at $z<0.45$, observed as part of the DESI DR1 survey. Using [O~III] and H$\alpha$ emission lines, we identified individual outflow candidates and quantified their mass-loading factors ($\eta$) and potential to enrich the surrounding CGM and/or IGM. We further compared these measurements with stacked spectra of the more broad low-mass population to assess how individually detected outflows relate to the average feedback properties of star-forming galaxies. Our main foundings are summarized as follows.
\begin{itemize}
\item We detect robust signature of ionized outflows in $\sim1\%$ of the overall dwarf galaxy parent sample, corresponding to 5781 sources. Most of these galaxies are classified as star-forming, with $\sim300$ of them being composite systems or AGNs. All of them occupy the starburst-dominated region of the stellar mass-SFR diagram, consistent with stronger emission lines and outflow-driven winds.
\item The derived outflow efficiencies span $\eta\sim0.001-1$, with a median value of $\eta\sim0.01$. These results confirm previous findings based on smaller samples of low-mass starbursts using similar ionized gas tracers. Higher mass-loading factors are typically found in galaxies hosting AGNs or in composite systems, consistent with trends observed in more massive local galaxies.
\item A large fraction of galaxies host high-velocity ionized gas capable of escaping the central galaxy and reaching the CGM, ranging from $\sim78\%$ of star-forming galaxies to $\sim88\%$ of composite systems and AGNs. In contrast, only $\sim11\%$ of star-forming galaxies reach velocities sufficient to escape the dark-matter halo, compared to $\sim56\%$ and $\sim68\%$ of composite systems and AGNs, respectively. 
This suggests that star-formation-driven outflows predominantly redistribute ionized gas into the CGM, while AGN activity can more efficiently accelerate gas to velocities capable of escaping the halo and potentially enriching the IGM.
\item Our measurements are broadly consistent with recent optical studies of local and high-redshift dwarfs, though the literature shows substantial scatter due to methodological differences and the specific gas phases probed. The ionized component likely represents only a minor fraction of the total outflow mass, in agreement with multi-phase studies of more massive systems.
\item We find that the mass-loading factor decreases with increasing stellar mass, qualitatively reproducing the behaviour predicted by cosmological simulations. However, the normalization of the observed relation remains up to two orders of magnitude below theoretical expectations. The discrepancy likely reflects a combination of the different gas phases probed by observations and simulations, together with systematic uncertainties in observational estimates of the mass-loading factor, rather than necessarily indicating fundamentally different feedback efficiencies.
\item Comparing individual outflow detections with stacked spectra of the star-forming low-mass population reveals systematically lower average mass-loading factors for the stacked measurements. However, this difference progressively decreases toward larger offsets from the star-forming main sequence ($\Delta$MS), where stacked galaxies approach the locus occupied by individual detections. The positive correlation observed between $\eta$ and specific star-formation rate suggests that elevated star-formation activity and burstiness are key regulators of feedback efficiency. Together with the steepening of the $\eta-M_\star$ relation observed in stacked starburst populations, these results support a scenario in which the dominant driving mode of galactic winds transitions from momentum-driven in average star-forming galaxies toward increasingly energy-driven feedback in burstier systems.
\end{itemize}
Overall, our results demonstrate that ionized outflows constitute an important feedback channel in dwarf galaxies, although they likely trace only a limited fraction of the total mass and energy carried by galactic winds. Individually detected outflows appear to preferentially probe phases of enhanced star formation, while stacked spectra provide a more representative view of the average feedback conditions across the broader galaxy population. In the future, we will extend this analysis to low-mass, high-redshift galaxies to explore any possible evolution of feedback across cosmic time. Finally, we stress the importance of obtaining a more complete, multi-phase picture of outflows. Observations of atomic and molecular gas in the low-mass regime using e.g., single-dish facilities such as APEX or IRAM 30m will provide essential constraints on the total outflow mass, energetics, and the interplay between different gas phases, helping to clarify the role of feedback in shaping galaxy evolution. 


\begin{acknowledgements}
D.D acknowledges support from the Polish National Agency for Academic Exchange (Bekker grant BPN /BEK/2024/1/00029/DEC/1). This research used data obtained with the Dark Energy Spectroscopic Instrument (DESI). DESI construction and operations is managed by the Lawrence Berkeley National Laboratory. This material is based upon work supported by the U.S. Department of Energy, Office of Science, Office of High-Energy Physics, under Contract No. DE–AC02–05CH11231, and by the National Energy Research Scientific Computing Center, a DOE Office of Science User Facility under the same contract. Additional support for DESI was provided by the U.S. National Science Foundation (NSF), Division of Astronomical Sciences under Contract No. AST-0950945 to the NSF’s National Optical-Infrared Astronomy Research Laboratory; the Science and Technology Facilities Council of the United Kingdom; the Gordon and Betty Moore Foundation; the Heising-Simons Foundation; the French Alternative Energies and Atomic Energy Commission (CEA); the National Council of Humanities, Science and Technology of Mexico (CONAHCYT); the Ministry of Science and Innovation of Spain (MICINN), and by the DESI Member Institutions: www.desi.lbl.gov/collaborating-institutions. The DESI collaboration is honored to be permitted to conduct scientific research on I’oligam Du’ag (Kitt Peak), a mountain with particular significance to the Tohono O’odham Nation. Any opinions, findings, and conclusions or recommendations expressed in this material are those of the author(s) and do not necessarily reflect the views of the U.S. National Science Foundation, the U.S. Department of Energy, or any of the listed funding agencies. This research uses services or data provided by the SPectra Analysis and Retrievable Catalog Lab (SPARCL) and the Astro Data Lab, which are both part of the Community Science and Data Center (CSDC) program at NSF National Optical-Infrared Astronomy Research Laboratory. NOIRLab is operated by the Association of Universities for Research in Astronomy (AURA), Inc. under a cooperative agreement with the National Science Foundation.
\end{acknowledgements}

%
%

\bibliographystyle{aa} 
\bibliography{aanda.bib} 

\begin{appendix}

\section{Representative examples of spectral decomposition}\label{app:spectral_fits}
\begin{figure*}[t!]
    \begin{center}
	\includegraphics[width=0.75\textwidth,height=1.1\textwidth]{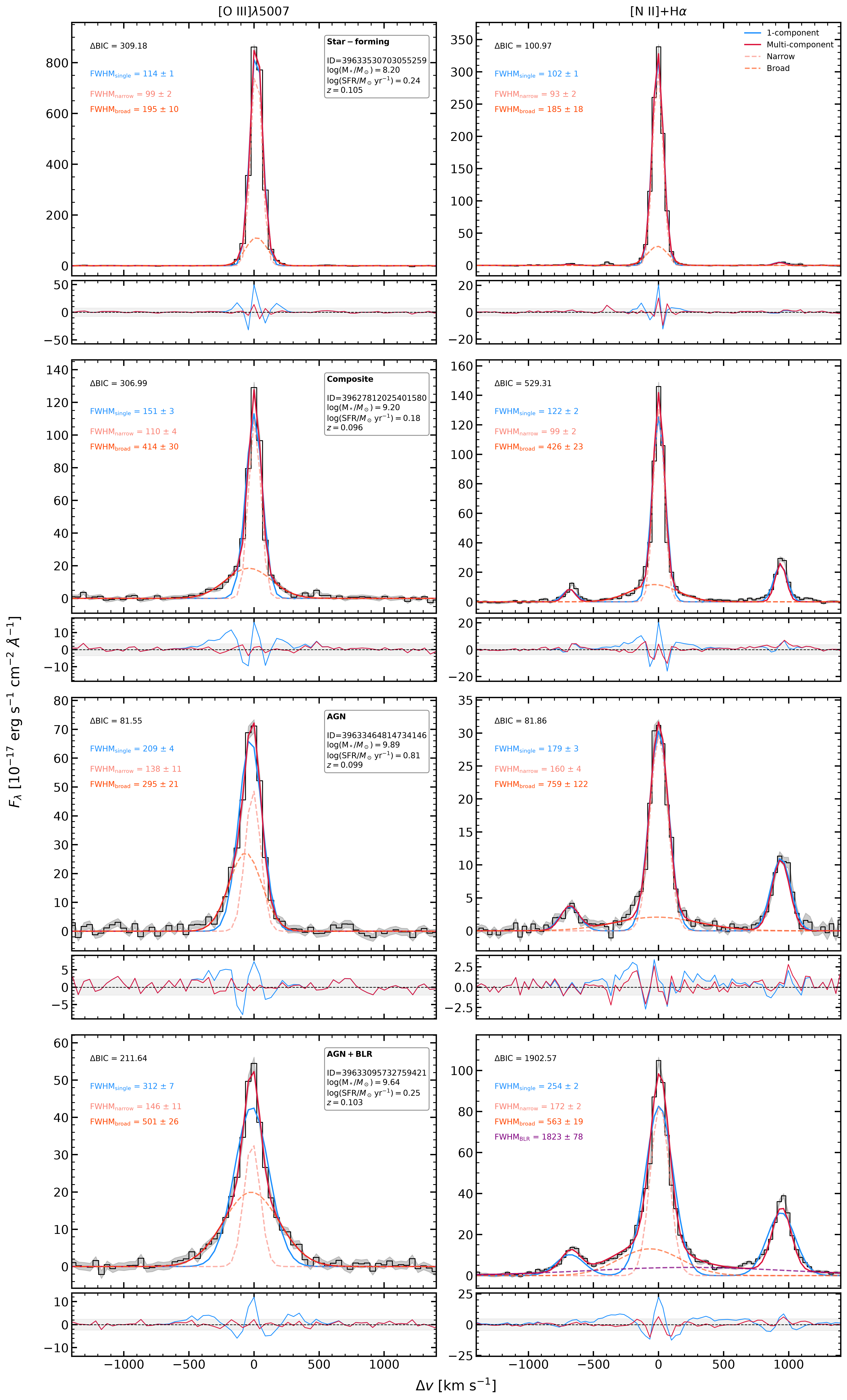}
	\end{center}
    \caption{Examples of spectral fitting for [O\,III]$\lambda5007$ (left) and [N\,II]+H$\alpha$ (right) emission lines for four representative galaxies in our outflow sample. The galaxies are ordered by increasing stellar mass from top to bottom and comprise a star-forming galaxy, a composite system, an AGN host, and an AGN for which an additional very broad H$\alpha$ component is preferred. For each galaxy, the continuum-subtracted spectrum is shown as a black histogram, with the corresponding uncertainty indicated by the grey shaded region. Single- and multi-component fits are shown by the blue and red solid lines, respectively, while the narrow and broad components of the multi-component fit are shown by the red dotted and dashed lines. In the bottom-right panel, the purple dashed line indicates the additional very broad H$\alpha$ component associated with BLR emission. The FWHM of the fitted components (in km/s, uncorrected for instrumental broadening) and the corresponding $\Delta$BIC are reported in each panel. The bottom panels show the residuals of the single- and multi-component fits. Source classification, DESI TARGETID, stellar mass, SFR, and redshift are reported in the left panels.}
    \label{fig:additional_fits}
\end{figure*}

Figure~\ref{fig:additional_fits} presents additional examples of the spectral decomposition described in Sect.~\ref{subsec:fitting}. The galaxies were selected to illustrate the performance of our fitting procedure across the stellar-mass range and different BPT classifications spanned by our outflow sample. We show examples of a star-forming galaxy, a composite system, and an AGN host, together with an additional AGN for which a very broad H$\alpha$ component is statistically preferred. For each galaxy, both the [O\,III] and [N\,II]+H$\alpha$ spectral regions are shown. The latter example illustrates how the additional H$\alpha$ component included in our fitting procedure accounts for possible BLR emission without attributing it to the intermediate-width component associated with the outflow.

\section{Comparison between outflow properties derived from [O~III] and H$\alpha$ lines}\label{app:oiii_ha_comparison}
\begin{figure*}[t!]
    \begin{center}
	\includegraphics[width=0.8\textwidth]{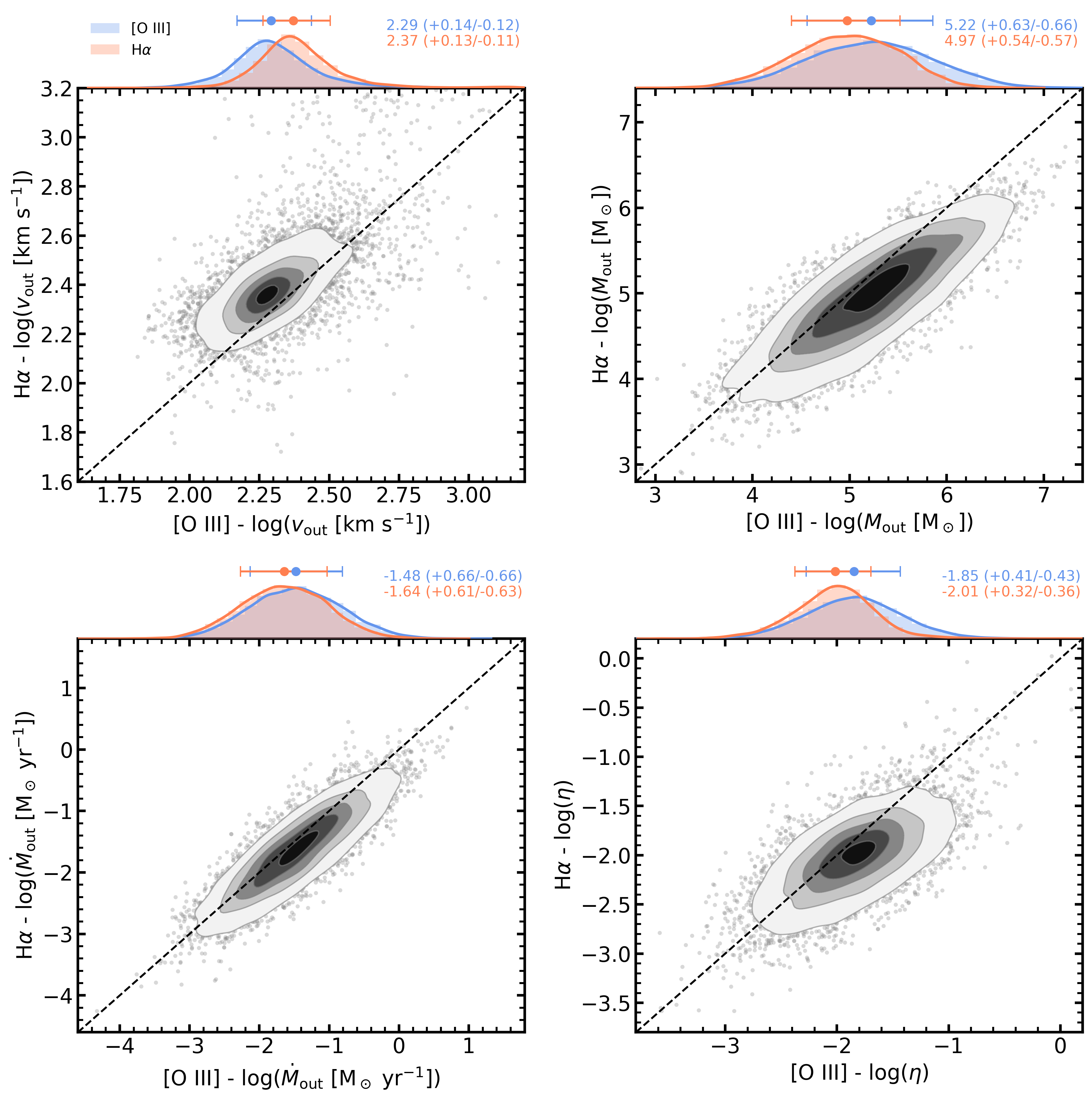}
	\end{center}
    \caption{Comparison of outflow properties derived from the [O~III] (light blue) and H$\alpha$ (coral) emission lines. From top left to bottom right, we show outflow velocity, mass, outflow rate, and mass-loading factor. In each main plot, contours start at 10\% density levels, increasing in steps of 20\%, while the dashed black line marks the 1:1 relation. The upper panels display the associated distributions, with circles and errorbars denoting the median and the 16th/84th percentiles, respectively. The corresponding numerical values are reported in the top-right corner.}
    \label{fig:oiii_ha_comp}
\end{figure*}

We present in Fig.~\ref{fig:oiii_ha_comp} a comparison of the ionized outflow properties derived from the [O~III] and H$\alpha$ emission lines, including outflow velocities, masses, mass outflow rates, and mass-loading factors. While individual measurements show significant scatter, the bulk of the data cluster around the 1:1 relation in all cases. The largest deviations are found for outflow velocities, where H$\alpha$ in particular may be affected by contamination from nearby [N~II] lines and BLR. Nevertheless, the overall distributions of all presented parameters are similar, with median values comparable within the uncertainties. This indicates that, despite object-to-object variations, both lines are statistically comparable and trace similar properties of the outflows.

\end{appendix}

\end{document}